\documentclass[%
 reprint,
 amsmath,amssymb,
 aps,
]{revtex4-1}

\usepackage{graphicx}% Include figure files
\usepackage{dcolumn}% Align table columns on decimal point
\usepackage{bm}% bold math
\begin{document}

%\preprint{APS/123-QED}

\title{Relaxation of Radiation-Driven Two-Level Systems Interacting \\ with a Bose-Einstein Condensate Bath}
%\title{Relaxation of two-level system Rabi oscillations \\ interacting with Bose-condensed thermostat}
%\thanks{A footnote to the article title}%

\author{Vadim M. Kovalev$^{1,2}$}
\email{vadimkovalev@isp.nsc.ru}
% \altaffiliation[Also at ]{Physics Department, XYZ University.}%Lines break automatically or can be forced with \\
\author{Wang-Kong Tse$^{3,4}$}%
\email{wktse@ua.edu}
% \email{Second.Author@institution.edu}
\affiliation{%
 $^1$Institute of Semiconductor Physics, Siberian Branch of Russian Academy of Sciences, Novosibirsk 630090, Russia\\
 $^2$Department of Applied and Theoretical Physics, Novosibirsk State Technical University, Novosibirsk 630073, Russia\\
  $^3$Department of Physics and Astronomy, The University of Alabama, Alabama 35487, USA\\
$^4$Center for Materials for Information Technology, The University of Alabama, Alabama 35401, USA}

%\collaboration{MUSO Collaboration}%\noaffiliation

\date{\today}% It is always \today, today,
             %  but any date may be explicitly specified

\begin{abstract}

We develop a microscopic theory for the relaxation dynamics of an optically pumped two-level
system (TLS) coupled to a bath of weakly interacting Bose gas. Using
Keldysh formalism and diagrammatic perturbation theory, 
%non-equilibrium Green's functions, 
expressions for the relaxation times of the TLS Rabi oscillations
are derived when the boson bath is in the normal state and the Bose-Einstein condensate (BEC) state. 
%We obtain expressions for the
%relaxation times of the TLS Rabi oscillations in the normal state and the
%Bose-Einstein condensate (BEC) state regimes of the bosonic bath. 
%The results we have obtained for the relaxation times of the TLS Rabi
%oscillations shows that a phase transition of the bosonic bath from the normal to
%the Bose-Einstein condensate (BEC) state strongly influences the
%relaxation process. 
We apply our general theory to consider an irradiated quantum dot
coupled with a boson bath consisting of a two-dimensional dipolar
exciton gas. When the bath is in the BEC regime, relaxation of the
Rabi oscillations is due to both condensate and non-condensate
fractions of the bath bosons for weak TLS-light coupling and
dominantly due to the non-condensate fraction for strong TLS-light
coupling.  
%is mainly due to coupling between the TLS and the non-condensate bath bosons. 
%We apply our general theory to consider an  irradiated quantum dot coupled with a two-dimensional dipolar exciton
%gas.  In particular, when the bath is in
%the BEC regime, relaxation of the Rabi oscillations is mainly due to
%coupling between the TLS and the non-condensate bath bosons.  
Our theory also shows that a phase transition of the bath from the normal to
the BEC state strongly influences the relaxation rate of the TLS Rabi oscillations. 
%Our results show that, While 
The TLS relaxation rate is approximately independent of the pump
field frequency and monotonically dependent on the field strength 
when the bath is in the low-temperature regime of
the normal phase. Phase transition of the dipolar exciton gas leads to a non-monotonic dependence of the
TLS relaxation rate on both the pump field frequency and field
strength, providing a characteristic signature for the detection of BEC phase transition of the coupled dipolar exciton gas. 
\end{abstract}

%\pacs{Valid PACS appear here}% PACS, the Physics and Astronomy
                             % Classification Scheme.
%\keywords{Suggested keywords}%Use showkeys class option if keyword
                              %display desired
\maketitle

%\tableofcontents

\section{\label{sec:level1}Introduction}

% One of the simplest and most ubiquitous models in quantum
% physics is the single-mode spin-boson model, consisting
% of a two-level system coupled to a quantum harmonic
% oscillator. In quantum optics, it describes an atom coupled
% to an electromagnetic field mode [1,2]; in condensed matter
% physics, it lies at the heart of the Holstein model for
% electrons coupled to phonon modes of a crystal lattice [3].
% More recently, implementations of this model have been
% achieved in superconductor [4–7] and semiconductor [8]
% systems. Still other proposals have involved mechanical
% oscillators [9,10]. Although the model itself is quite simple,
% it displays a rich variety of behaviors, encapsulating
% many of the unique aspects of quantum theory.

The dynamics of a quantum two-level system (TLS) is a topic
of fundamental importance. Its sustained influence % relevance
 is evident in the continual interest in the dynamics of spin or pseudospin systems ranging
from quantum optics \cite{QO_ref_1,QO_ref_2} to quantum information
\cite{QI_ref_1,QI_ref_2,QI_ref_3}. The 
spin-boson model ~\cite{2_Weiss} captures the interaction between the TLS and its
environment by a spin $1/2$ degree of freedom coupled linearly to an oscillator
bath ~\cite{3_Makhlin}. Despite the simplicity of such a model, 
it exhibits a rich variety of behavior and describes a diverse array of physical systems and phenomena ~\cite{1_Leggett,2_Weiss}.
%it exhibits immensely rich physics and successfully 
%describes a large number of different physical systems and phenomena ~\cite{1_Leggett},~\cite{2_Weiss}.
One of the quantum systems that is well described by a TLS is the
quantum dot (QD). Fueled by interests in quantum information
processing, coherent optical control of quantum dots has seen
substantial development in the past decade \cite{QD_review1}. New 
functionalities or tuning capabilities can be achieved with hybrid systems by further
coupling QDs to other materials, such as nano-sized cavity \cite{QD_cavity}, graphene \cite{QD_graphene}, and superconductor \cite{QD_SC}.

Hybrid quantum systems comprising a fermion gas coupled to a boson gas
%have been extensively studied and 
constitute the condensed
matter analogue of $^3$He-$^4$He mixtures.  
%(where $^3$He is a fermion and $^4$He a boson) widely-studied in low-temperature physics.
In systems where an electron gas is coupled with excitons or
exciton-polaritons, it was recently predicted that the 
transition of the excitonic subsystem to the Bose Einstein condensate
(BEC) phase strongly modifies the properties of the electronic subsystem, resulting
in polariton-mediated superconductivity and 
supersolidity~\cite{6_RefImamogluPRB2016,7_RefShelykhPRL2010,8_RefShelykhPRL1051404022010,8_1Shelykh2012,9_RefMatuszewskiPRL1080604012012}.
The topic of phase transition of the exciton or exciton-polariton Bose
system into the BEC state is itself an intriguing topic that has
garnered much attention
~\cite{4_1Butov,4_2Butov,4_3Butov,5_Timofeev,5_1Timofeev}. For
dipolar exciton systems realized in GaAs double quantum well (DQW)
structures, the critical temperature to reach the condensate phase is
about $3-5\,\mathrm{K}$. Recent works have demonstrated that
double-layer structures based on transition metal dichalcogenides
(TMD) monolayers ~\cite{10_3butovAPL2016,10_4Rivera2015} can further push the transition temperature to $\sim
10-30\,\mathrm{K}$ ~\cite{10_2butovNature2014,10_5Wu2015,10_6Berman2016}. 

Motivated by recent interest in hybrid fermion-boson systems and
exciton-polariton physics mentioned above, in this paper we 
consider a radiation-driven quantum dot coupled to a dipolar exciton gas and study the influence of
the latter's BEC phase transition on the dynamics and relaxation of
the QD states. The problem of TLS relaxation coupled to a fermionic bath transitioning to a superconducting state
was studied in the context of metallic glasses ~\cite{10_6.1BlackFulde1979}; however, the
question of TLS relaxation coupled to a \textit{bosonic} bath
transitioning to a BEC state has not, up to the authors' knowledge,
been considered before. 
% We note that the problem of TLS 
% relaxation coupled to a fermionic bath transitioning to a superconducting state
% was studied before in the context of metallic glasses ~\cite{10_6.1BlackFulde1979}. However, the
% question of TLS relaxation coupled to a bosonic bath transitioning to
% a BEC state has not, up to the authors' knowledge, been considered
% before. 
It is noteworthy to mention that our current work is closely connected to the 
problem of a mobile impurity moving in a BEC 
\cite{11_Mobile_impurity}, since the renormalization of physical properties of a moving electron that strongly interacts
with the surrounding medium (polaron problem) can be described by a quantum particle coupled with a bath ~\cite{2_Weiss}.
% It is also worth noting that our current problem is connected to the
% problem of a mobile impurity moving in a BEC \cite{11_Mobile_impurity}, which in turn is reminescent of the polaron
% problem. The renormalization of physical properties of a moving electron that strongly interacts
% with the surrounding medium (polaron problem) can be described by a quantum particle coupled with a bath ~\cite{2_Weiss}.

Our theory consists of a TLS modeling the ground and
lowest excited states of the QD, which is coupled
~\cite{10_9.1Melik1978,10_9.2Weisskopf1930,10_9.3Gordon1963}
to a bath of weakly-interacting Bose
gas modeling the dipolar exciton system. In contrast to the simple spin-boson model, the
interaction between the QD and the 2D dipolar exciton gas in our system is
described by a \emph{nonlinear} coupling Hamiltonian. We take the 
Bose gas to be weakly interacting, exhibiting a normal phase as
well as a BEC phase described by the Bogoliubov model~\cite{10_10Beliaev1958,10_11PitaevskiiStringari2003,10_10.1Griffin1998}.
Our results demonstrate that the damping of the Rabi oscillations of the TLS is
highly sensitive to the phase transition of the bosonic bath.

The rest of our paper is 
organized as follows. The second section is devoted to the development
of general theory for the relaxation rate of an illuminated TLS
coupled to a bosonic bath. We then apply our general results to the situation of a QD coupled with a dipolar exciton
gas in the third section. Finally, in the fourth section
we present numerical results of the relaxation rates and discuss their 
behavior as a function of the optical pump field's parameters. % and present our conclusions in fifth section. 
In the Appendix we present details of our calculations. 

\section{General theory}
%\subsection{Isolated TLS and Rabi oscillations}
\subsection{Driven TLS and Rabi oscillations}
First, we consider dynamics of isolated TLS system under strong
external electromagnetic field and describe the system's response
using the
%pure Rabi-oscillation effect using
non-equilibrium Keldysh Green function technique. The TLS Hamiltonian
is given by
\begin{equation}\label{eq1}
%\begin{bmatrix}
\hat{\tilde{H}}_0(t)=\left(
          \begin{array}{cc}
            \Delta & \lambda e^{-i\omega t} \\
            \lambda^\ast e^{i\omega t} & -\Delta \\
          \end{array}
        \right),
\end{equation}
where $\pm\Delta$ are the energies of the upper and lower states of
the TLS, and quantities with an overhead caret ($\,\hat{}\,$) symbol denotes a
matrix quantity. The interaction Hamiltonian with the electromagnetic field is
written here in the Rotating Wave Approximation (RWA). $\lambda$ is
the interaction matrix element and $\omega$ the frequency of the
electromagnetic field. It is also assumed in Eq.~(\ref{eq1}) that the
wavelength of the electromagnetic field is much larger than the
geometrical size of the TLS so that the field is uniform on our scale
of interest. In this work, we denote quantities in the laboratory frame and the rotating frame, respectively, with and without an
overhead tilde. %In the rest of this paper, all quantities are expressed in the rotating frame.

%Without interaction with an external bosonic bath,
The dynamics of the TLS is described by the time-ordered Green's function $\hat{\tilde{G}}_0(t,t')$
satisfying the equation of motion
%differential equation
%
\begin{gather}\label{eq2}
\left(
  \begin{array}{cc}
    i\partial_t-\Delta & -\lambda e^{-i\omega t} \\
    -\lambda^\ast e^{i\omega t} & i\partial_t+\Delta \\
  \end{array}
\right)\hat{\tilde{G}}_0(t,t')=\delta(t-t').
\end{gather}
To remove the explicit time dependence,
%of the Hamiltonian Eq.~(\ref{eq1}),
it is convenient to transform this equation to the rotating frame
using the unitary transformation % operator
\begin{gather}\label{eq3}
\hat{S}=\left(
          \begin{array}{cc}
            e^{i\omega t/2} & 0 \\
            0 & e^{-i\omega t/2} \\
          \end{array}
        \right),\,\,\,\hat{S}^{-1}=\left(
          \begin{array}{cc}
            e^{-i\omega t/2} & 0 \\
            0 & e^{i\omega t/2} \\
          \end{array}
        \right).
\end{gather}
%
%The advantage of this transformation to the new reference frame is that it removes the explicit time dependence from the Hamiltonian.
As a result, the Green's function in the rotating frame,
$\hat{G}_0(t,t')=\hat{S}(t)\hat{\tilde{G}}_0(t,t')\hat{S}^{-1}(t')$, is described by the equation
\begin{gather}\label{eq4}
\left(
  \begin{array}{cc}
    i\partial_t-\varepsilon_0 & -\lambda  \\
    -\lambda^\ast  & i\partial_t+\varepsilon_0 \\
  \end{array}
\right)\hat{G}_0(t,t')=\delta(t-t'),
\end{gather}
where $\varepsilon_0=\Delta-{\omega}/{2}$.
%Using the solution of Eq.~(\ref{eq4}), it is possible to describe the Rabi oscillation phenomenon.
%To do this, we need first the retarded Green's function, derived in the Appendix:
%It is easy to find it from Eq.(\ref{eq4}). The simple calculation yields
To find the self-energies and lifetimes, we need the retarded and the
lesser components of the non-equilbrium Green's function. The
retarded Green's function is derived as (see Appendix A),
\begin{equation}
%\bar{G}^{R}_{ij}(\varepsilon)
%=\frac{A_{ij}}{\varepsilon-\Omega+i\delta}
% +\frac{B_{ij}}{\varepsilon+\Omega+i\delta},\\\nonumber
\hat{G}_0^{R}(\varepsilon)
=\frac{\hat{A}}{\varepsilon-\Omega+i\delta}
 +\frac{\hat{B}}{\varepsilon+\Omega+i\delta},\label{eq5}
\end{equation}
where $\Omega=\sqrt{\varepsilon_0^2+|\lambda|^2}$ is the Rabi
frequency,
%$u^2=\frac{1}{2}\left(1+\frac{\varepsilon_0}{\Omega}\right),\,\,\,
%v^2=\frac{1}{2}\left(1-\frac{\varepsilon_0}{\Omega}\right),\,\,\,(uv)^{(\ast)}=\frac{\lambda^{(\ast)}}{2\Omega}$,
and $\hat{A}$ and $\hat{B}$ are matrices defined as
\begin{equation}
\hat{A}=\left(
   \begin{array}{cc}
     u^2 & uv \\
     u^\ast v^\ast & v^2 \\
   \end{array}
 \right),\,\,
\hat{B}=\left(
                                               \begin{array}{cc}
                                                 v^2 & -uv \\
                                                 -u^\ast v^\ast & u^2 \\
                                               \end{array}
                                             \right),\label{eq5_1}
\end{equation}
with $u^2=\left(1+{\varepsilon_0}/{\Omega}\right)/2$,
$v^2=\left(1-{\varepsilon_0}/{\Omega}\right)/2$ and
$uv={\lambda}/{2\Omega}$. Eqs.~(\ref{eq5})-(\ref{eq5_1}) imply that new
quasiparticles emerge from the light-matter coupling that renormalizes the
original TLS states into dressed states with energies $\pm \Omega$. $\hat{A}$ and
$\hat{B}$ are the projection operators to these dressed states.

%To determine the Rabi oscillations of the TLS
%transition probability between energy level of TLS we apply the Feynman's ideology: the wave function of TLS at time $t$ couples with wave function at $t=0$ as
%
It is instructive to recover the result for Rabi oscillations using the
above retarded Green's function Eq.~(\ref{eq5}). The TLS wave function at a
latter time $t$ is obtained by propagating the initial time ($t = 0$)
wave function,
%can be propagated to yield the wave function at a latter time $t$:
%
\begin{eqnarray}
\psi_i(t) &=&[\hat{S}^{-1}(t)\hat{G}^R(t)\hat{S}(0)]_{ij}\psi_j(0)
              \nonumber \\
&=& \left[\hat{S}^{-1}(t)\int\frac{d\varepsilon}{2\pi}\hat{G}^R(\varepsilon)e^{-i\varepsilon
  t}\right]_{ij}\psi_j(0), \label{eq6}
\end{eqnarray}
where $\psi_i$ are the wave functions of the TLS upper $(i=1)$ and lower
$(i=2)$ levels. Here $\hat{S}^{-1}(t)\hat{G}^R(t)\hat{S}(0) \equiv
\hat{\tilde{G}}^R(t,0)$ is the retarded Green's function in the
laboratory frame. When only the lower level is initially occupied,
$\psi_2(0)=1$ and $\psi_1(0)=0$. Using Eqs.(\ref{eq3})-(\ref{eq5_1}),
the transition probability to the upper level is then given by %$|\langle\psi_2^+(0)\psi_1(t)\rangle|^2$.
%Straightforward calculations using Eqs.(\ref{eq5}) and (\ref{eq6})
%then yield
%
\begin{eqnarray}
|\langle\psi_2^+(0)\psi_1(t)\rangle|^2&=&\left|\frac{\lambda}{\Omega}\sin(\Omega
  t)e^{-i\omega t/2}\right|^2 \nonumber \\
&=&\frac{|\lambda|^2}{2\Omega^2}\left(1-\cos\Omega t\right), \label{eq7}
%|\langle\psi_2^+(0)\psi_1(t)\rangle|^2=\left|\frac{\lambda}{\Omega}\sin(\Omega t)e^{-i\omega t/2}\right|^2=\frac{|\lambda|^2}{2\Omega^2}\left(1-\cos\Omega t\right),
\end{eqnarray}
which is the Rabi oscillations \cite{12_LL}.
%This is exactly the textbook result \cite{1}.

The lesser Green's function can be expressed in terms of the
distribution functions $n_{\pm \Omega}$ of the upper and lower dressed states as
\begin{eqnarray}
\hat{G}^{<}(\varepsilon) &=&
                             -n_{\varepsilon}\left[\hat{G}^{R}(\varepsilon)-\hat{G}^{A}(\varepsilon)\right]
 \nonumber \\
&=& 2\pi i
    n_{\varepsilon}\left[\hat{A}\delta(\varepsilon-\Omega)+\hat{B}\delta(\varepsilon+\Omega)\right]
    \nonumber \\
&=& 2\pi i
    \left[\hat{A}n_{\Omega}\delta(\varepsilon-\Omega)+\hat{B}n_{-\Omega}\delta(\varepsilon+\Omega)\right]. \label{Glesser1}
\end{eqnarray}
Note that $n_{\Omega}+n_{-\Omega} = 1$. Assuming the radiation is turned on adiabatically, we can obtain
$n_{\pm \Omega}$ in the following. The density matrix $\hat{f}$ in the original basis of TLS upper and lower
levels satisfies the kinetic equation (see Appendix A):
\begin{eqnarray} \label{KE}
%\begin{equation}
&&\frac{\partial \hat{f}}{\partial t} +i[\hat{H}_0,\hat{f}] = 0, \label{KE} \\
&&\hat{f} = \left(\begin{array}{cc}
                                                 f_{11} & f_{12} \\
                                                 f_{21} & f_{22} \\
                                               \end{array}
                                             \right), \label{distfcn}
%\end{equation}
\end{eqnarray}
where the subscripts $1,2$ respectively denotes the original (\textit{i.e.},
unrenormalized by light) upper and lower levels of the
TLS, and $\hat{H}_0 = \hat{S}(t)\hat{\tilde{H}}_0(t,0)\hat{S}^{-1}(0)$ is the
Hamiltonian in the rotating frame. Writing the Hamiltonian as $\hat{H}_0 =
\hat{\bm{\sigma}}\cdot {\bm{B}}_0/2$, we can define an effective
magnetic field ${\bm{B}}_0 =2\lambda_R{\bm{e}_x}-2\lambda_I{\bm{e}_y}+
(2\Delta-\omega){\bm{e}_z}$
that drives the TLS pseudospin degrees of freedom, where
$\lambda_{R,I}$ are the real and imaginary parts of $\lambda$ and $\bm{e}_{x,y,z}$ the unit vectors along the $x,y,z$ directions.
Then, decomposing the density matrix $\hat{f}$ as $\hat{f} =
C+\bm{S}\cdot\hat{\bm{\sigma}}/2$, the
kinetic equation can be written as a Bloch equation:
\begin{equation}
\frac{\partial \bm{S}}{\partial t}+\bm{S}\times {\bm{B}}_0=
0. \label{Bloch}
\end{equation}
From the definition of $\hat{f}$ in Eq.~(\ref{distfcn}), we can relate the
distribution functions in the two representations as $S_{z} = f_{11}-f_{22}$ and
$S^{(+)} \equiv S_{x}+iS_{y} = 2f_{12}^* = 2f_{21}$. With the laser field switched on adiabatically, the
optical response follows adiabatically the driving field and is therefore stationary
in the rotating frame, \textit{i.e.}, $\partial/\partial t =
0$. Before laser is turned on, the TLS initial state is in the lower level so that $\bm{S}(t =
0) = -{\bm{e}_z}$. Since $\vert \bm{S} \vert$ is a constant of motion,
this implies that $\vert \bm{S}(t) \vert = 1$ for all times $t$. Here we focus on the regime without
population inversion, so that $S_{z} = f_{11}-f_{22}$ is always less
than zero. We obtain $\bm{S}$ as
\begin{eqnarray}
S^{(+)} &=& -\frac{\mathrm{sgn}(2\Delta-\omega)2\lambda^*}{\sqrt{(2\Delta-\omega)^2+4\lambda^2}}, \\
S_{z} &=& -\frac{\vert 2\Delta-\omega \vert}{\sqrt{(2\Delta-\omega)^2+4\lambda^2}}.
\end{eqnarray}
Using $f_{11}+f_{22} =1$, we also find the density matrix
$\hat{f}$ in the original basis of the TLS upper and lower levels:
\begin{eqnarray}
f_{11} &=& \frac{1}{2}\left[1-\frac{\vert 2\Delta-\omega
    \vert}{\sqrt{(2\Delta-\omega)^2+4\lambda^2}}\right], \label{fpp} \\
f_{12} &=&
-\frac{\mathrm{sgn}(2\Delta-\omega)2\lambda}{\sqrt{(2\Delta-\omega)^2+4\lambda^2}}. \label{fpm}
\end{eqnarray}
On the other hand, Eq.~(\ref{Glesser1}) gives the density matrix in the basis of the dressed quasiparticles as follows
\begin{eqnarray}
\hat{f}(t) &=& -i\hat{G}^{<}(t,t) =
  -i\int\frac{d\varepsilon}{2\pi}\hat{G}^{<}(\varepsilon), \nonumber \\
&=& \hat{A}n_{\Omega}+\hat{B}n_{-\Omega}.
\label{Glesser2}
\end{eqnarray}
We can determine $n_{\Omega}$ by comparing the expressions of the
density matrix in Eq.~(\ref{Glesser2}) and Eq.~(\ref{fpp}). The $11$ element, for instance, gives $f_{11} = u^2
n_{\Omega}+v^2(1-n_{\Omega})$ or $n_{\Omega} =
({f_{11}-v^2})/({u^2-v^2})$, from which we determine
\begin{equation}
%&=& %\frac{f_{11}-v^2}{u^2-v^2} \nonumber \\
n_{\pm \Omega} = \frac{1}{2}\left[1\mp \mathrm{sgn}\left(\varepsilon_0\right)\right]. \label{nOmega}
%    \frac{1}{2}\left[1-\mathrm{sgn}\left(\Delta-\omega/2\right)\right]. \label{nOmega}
\end{equation}
It follows that $n_{\Omega}$ takes on the values $0$ or $1$ depending on
whether the light frequency $\omega$ is smaller or larger, respectively, than the energy
difference $2\Delta$ of the TLS.

\subsection{Coupling to bosonic bath and TLS self-energies}
In the absence of bath coupling, the TLS is described by bare Green's
functions with a vanishing level broadening. To %highlight 
focus on the effects of bosonic bath coupling, 
we ignore the effects of spontaneous and stimulated emission due to electrons' coupling to light. 
%This means that the TLS is described by Green's functions without any disorder broadening. 
The only damping effects on the TLS dynamics, once coupled to the
bosonic bath, will be due to interlevel transitions caused by absorption or emission of the
bosons.
To analyze the TLS-bath coupling, we add to the bare TLS
Hamiltonian Eq.~(\ref{eq1}) the TLS interaction term with the bath and
the bath Hamiltonian
\begin{gather}\label{eq8}
\left(
                  \begin{array}{cc}
                    W_{11}[\varphi] & 0 \\
                    0 & W_{22}[\varphi] \\
                  \end{array}
                \right)+\hat{H}_{bath}[\varphi].
\end{gather}
Here the first term is TLS-bath coupling Hamiltonian %written in
                                %matrix form,
where the matrix elements describe the interaction
of the upper and lower levels with the bath bosons. Both terms in
Eq.~(\ref{eq8}) are functionals of the quantum field $\varphi$, which
describes the dynamics of bath degrees of freedom. The structure of
the bath Hamiltonian depends on whether it is in the normal or
Bose-condensed phase and will be specified later on.
We assume short-range interaction between the TLS and the bosonic
bath
%We will assume that the interacting part of Eq.(\ref{eq8}) has the following 'contact-like' structure
%
\begin{gather}\label{eq9}
W_{ii}[\varphi]=g_i\int d\textbf{r}|\psi_{i}(\textbf{r})|^2|\varphi(\textbf{r},t)|^2,
\end{gather}
with the coupling constant $g_i$. The form of this
interaction contains $\varphi^2$ and is markedly different from the
conventional coupling to a phonon-like bath, which is linear in
$\varphi$~\cite{2_Weiss}. Anticipating further application of our general theory to
consider a QD interacting with a 2D exciton gas, we note that $W_{ii}$
in Eq.~(\ref{eq9}) takes into account  the most important direct contribution to the Coulomb interaction energy between electrons in the QD and
the 2D exciton gas.

To elucidate the influence of the bosonic bath on the TLS dynamics, we
use the diagrammatic perturbation theory. Within this approach the
retarded Green function can be found from the Dyson equation, %having in our case the matrix form
\begin{equation}\label{eq10}
\hat{G}^R=\hat{G}_0^{R}+\hat{G}_0^{R}\hat{\Sigma}^R\hat{G}^{R},
%\bar{G}^R_{ij}=\bar{G}^{0R}_{ij}+\bar{G}^{0R}_{im}\bar{\Sigma}^R_{mn}\bar{G}^{R}_{nj},
\end{equation}
where the bare Green's function $\hat{G}_0^{R}$ is given by Eq.~(\ref{eq5}). The solution of Eq.~(\ref{eq10}) gives the
interacting Green's function as follows
\begin{equation}
\hat{G}^R =
\frac{1}{\Lambda}\left[{\hat{G}_{0}^{R}-\det(\hat{G}_0^{R})\sigma_y
  (\hat{\Sigma}^R)^{\mathrm{T}}\sigma_y}\right],  \label{eq11}
\end{equation}
%
%
%\hat{G}^R =
%\frac{\hat{G}_{0}^{R}-\det(\hat{G}_0^{R})\sigma_x
%  (\hat{\Sigma}^R)^{\mathrm{T}}\sigma_x}{ 1-\mathrm{Tr}(\hat{G}_0^{R}\hat{\Sigma}^R)+\det(\hat{G}_0^{R}\hat{\Sigma}^R)},  \label{eq11}
% \hat{G}^R = \frac{1}{\Lambda}\left(
%                  \begin{array}{cc}
%                    {G}_{0,11}^{R}-\det(\hat{G}_0^{R}){\Sigma}_{22}^R  & {G}_{0,12}^{R}-\det(\hat{G}_0^{R}){\Sigma}_{12}^R \\
%                    {G}_{0,21}^{R}-\det(\hat{G}_0^{R}){\Sigma}_{21}^R & {G}_{0.22}^{R}-\det(\hat{G}_0^{R}){\Sigma}_{11}^R \\
%                  \end{array}
%                \right),  \label{eq11}
%               \Lambda &=&
%                           1-Tr(\hat{G}_0^{R}\hat{\Sigma}^R)+\det(\hat{G}_0^{R}\hat{\Sigma}^R). \label{eq11}
%\bar{G}^R=\frac{1}{\Lambda}\left(
%                 \begin{array}{cc}
%                   \bar{G}_{0,11}^{R}-\det(\bar{G}_0^{R})\bar{\Sigma}_{22}^R  & \bar{G}_{0,12}^{R}-\det(\bar{G}_0^{R})\bar{\Sigma}_{12}^R \\
%                   \bar{G}_{0,21}^{R}-\det(\bar{G}_0^{R})\bar{\Sigma}_{21}^R & \bar{G}_{0.22}^{R}-\det(\bar{G}_0^{R})\bar{\Sigma}_{11}^R \\
%                 \end{array}
%               \right)\\
%               \Lambda=1-Tr(\bar{G}_0^{R}\bar{\Sigma}^R)+\det(\bar{G}_0^{R}\bar{\Sigma}^R).
where $\Lambda =
1-\mathrm{Tr}(\hat{G}_0^{R}\hat{\Sigma}^R)+\det(\hat{G}_0^{R}\hat{\Sigma}^R)$, $(\hat{\Sigma}^R)^{\mathrm{T}}$ is the matrix transpose of
$\hat{\Sigma}^R$, and $\sigma_y$ is the Pauli matrix. 
%It can be easily evaluated from Eqs.~(\ref{eq5})-(\ref{eq5_1}) that $\det(\hat{G}_0^{R}) = 1/(\varepsilon^2-\Omega^2+i\delta)$. 
% and $\det(\hat{G}_0^{R}\hat{\Sigma}^R)=\det(\hat{G}_0^{R})\det(\hat{\Sigma}^R)=0$.
%It can be easily seen from Eq.(\ref{eq5}) that $\det(\hat{G}_0^{R})=0$ and $\det(\hat{G}_0^{R}\hat{\Sigma}^R)=\det(\hat{G}_0^{R})\det(\hat{\Sigma}^R)=0$.
%To leading order in the coupling strength $\sim g_{11,22}^2, g_{11}g_{12}$, 
We arrive at the following form of the Green's
function (see Appendix B) 
%can be presented in a more compact form 
%
\begin{eqnarray}
\hat{G}^R(\epsilon) &\approx& 
%\frac{\hat{G}_0^{R}(\epsilon)}{1-\mathrm{Tr}\left[\hat{G}_0^{R}(\epsilon)\hat{\Sigma}^R(\epsilon)\right]}
%  \nonumber \\ &=& 
\frac{\hat{A}}{\epsilon-\Omega+i/2\tau_u}
+\frac{\hat{B}}{\epsilon+\Omega+i/2\tau_l}, \label{eq12}
%\bar{G}^R_{ij}(\epsilon)=\frac{\bar{G}^{0R}_{ij}(\epsilon)}{1-Tr\left[\bar{G}^{0R}(\epsilon)\bar{\Sigma}^R(\epsilon)\right]}\approx\\
%\approx\frac{A_{ij}}{\epsilon-\Omega+i/2\tau_u}
%+\frac{B_{ij}}{\epsilon+\Omega+i/2\tau_l},
\end{eqnarray}
where we have introduced the relaxation times $\tau_{u,l}$ of the upper (subscript $u$) and
lower (subscript $l$) dressed quasiparticles
\newpage
\begin{eqnarray}
\frac{1}{2\tau_u} &=&-\textmd{Im} \,
\mathrm{Tr}\left[\hat{A}\hat{\Sigma}^R(\Omega)\right],\label{eq13_1} \\
\frac{1}{2\tau_l} &=&-\textmd{Im} \,
                      \mathrm{Tr}\left[\hat{B}\hat{\Sigma}^R(-\Omega)\right], \label{eq13_2}
\end{eqnarray}
and neglected the shift of the levels due to the real part of
self-energy. To obtain expressions of the relaxation times, we apply
the diagrammatic perturbation theory in the leading order of %respect to
the TLS-bath interaction potential and account
for the lowest-order non-vanishing diagrams for the self-energy. The
explicit form of such diagrams depends on whether the bath is in the
normal or the condensate state. First we consider the normal state.

\subsubsection{Bath in the normal state}

\begin{figure}[!tb]
%\centerline{\input epsf \epsfysize=2cm \epsfbox{figure1.eps}}
\includegraphics[width=4.5cm,angle=0]{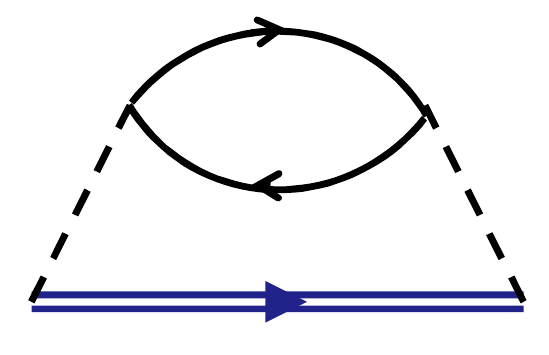}
\caption{Feynman diagram for TLS self-energy when the bath is
    in the normal phase. Double blue line represents the TLS Green function, dashed
  lines the TLS electron-bath interaction, and solid black lines
  the Green functions of the normal-state bath particles.}
%Feynman diagram for TLS self-energy in case of bath normal phase. Double blue line represents the TLS Green function, dashed lines are QD electron-hole pair -- bath interaction, solid black lines are Green functions of bath particles.}
\label{Figure1}
\end{figure}

In the normal state of the bath, we assume that the bosons are
non-interacting with a kinetic energy
$E_{\textbf{p}}=\vert\textbf{p}\vert^2/2m \equiv p^2/2m$
%\textbf{p}^2/2M$
and chemical potential $\mu$. To lowest order in the TLS-bath
interaction, the self-energy diagram is shown in Fig.~\ref{Figure1}. Detailed
calculation of the self-energy is presented in the Appendix C.
%Analytical expression of this diagram is calculated in the Appendix.
%Supplementary Material [?].
The result reads
\begin{widetext}
\begin{eqnarray}\label{eq14}
\hat{\Sigma}^R(\omega) &=& \sum_{\textbf{k},\textbf{p}}\left[
\hat{M}_{\textbf{k}}\hat{A}\hat{M}_{-\textbf{k}}\frac{(1-n_\Omega)[n_B(\xi_{\textbf{p}})-n_B(\xi_{\textbf{p}+\textbf{k}})]+n_B(\xi_{\textbf{p}+\textbf{k}})
  [1+n_B(\xi_{\textbf{p}})]}{\omega-\Omega+E_{\textbf{p}}-E_{\textbf{p}+\textbf{k}}+i\delta}\right.
\nonumber \\
&&\left.+\hat{M}_{\textbf{k}}\hat{B}\hat{M}_{-\textbf{k}}\frac{(1-n_{-\Omega}) [n_B(\xi_{\textbf{p}})-n_B(\xi_{\textbf{p}+\textbf{k}})]+n_B(\xi_{\textbf{p}+\textbf{k}}) [1+n_B(\xi_{\textbf{p}})]}{\omega+\Omega+E_{\textbf{p}}-E_{\textbf{p}+\textbf{k}}+i\delta}
\right],
\end{eqnarray}
\end{widetext}
%
% \begin{widetext}
% \begin{gather}\label{eq14}
% \hat{\Sigma}^R(\omega)=\sum_{\textbf{k},\textbf{p}}\left[
% \hat{M}_{\textbf{k}}\hat{A}\hat{M}_{-\textbf{k}}\frac{(1-n_\Omega)(n_{\textbf{p}}-n_{\textbf{p}+\textbf{k}})+n_{\textbf{p}+\textbf{k}}(1+n_{\textbf{p}})}{\omega-\Omega+E_{\textbf{p}}-E_{\textbf{p}+\textbf{k}}+i\delta}+
% \hat{M}_{\textbf{k}}\hat{B}\hat{M}_{-\textbf{k}}\frac{(1-n_{-\Omega})(n_{\textbf{p}}-n_{\textbf{p}+\textbf{k}})+n_{\textbf{p}+\textbf{k}}(1+n_{\textbf{p}})}{\omega+\Omega+E_{\textbf{p}}-E_{\textbf{p}+\textbf{k}}+i\delta}
% \right],
% \end{gather}
% \end{widetext}
%
where $\xi_{\textbf{p}}  = E_{\textbf{p}}-\mu$ is the energy of the bath
bosons rendered from the chemical potential, $n_B(\xi) =
[\exp(\xi/T)-1]^{-1}$ is the Bose distribution function, and
\begin{equation}
\hat{M}_{\textbf{k}}=\left(
                      \begin{array}{cc}
                        g_1\int d\textbf{r}e^{i\textbf{kr}}|\psi_1(\textbf{r})|^2 & 0 \\
                        0 & g_2\int d\textbf{r}e^{i\textbf{kr}}|\psi_2(\textbf{r})|^2 \\
                      \end{array}
                    \right),\label{eq15}
\end{equation}
%and $n_{\Omega}=1-n_{-\Omega}$ is the occupation number of the
%light-driven TLS.
Taking the imaginary part of Eq.~(\ref{eq14}), integrating over
the angle between $\textbf{k}$ and $\textbf{p}$, and
substituting the resulting expression into Eqs.~(\ref{eq13_1})-(\ref{eq13_2}) we find the relaxation times
%
%\begin{gather}\label{eq15.1}
\begin{eqnarray}
\frac{1}{2\tau_u} &=&\frac{m}{(2\pi)^2}\int_0^\infty
dk \left[\int_{k/2}^\infty
\frac{\alpha_kf_\Omega^u(p)pdp}{\sqrt{p^2-(k/2)^2}}\right.\label{eq15_1} \\
&&\left.+\int_{|{k}/2-{2m\Omega}/{k}|}^\infty
\frac{\beta_kf_{-\Omega}^u(p)pdp}{\sqrt{p^2-\left({k}/{2}-{2m\Omega}/{k}\right)^2}}\right],\nonumber\\
\frac{1}{2\tau_l} &=&\frac{m}{(2\pi)^2}\int_0^\infty
dk\left[\int_{k/2}^\infty
  \frac{\gamma_kf_{-\Omega}^l(p)pdp}{\sqrt{p^2-(k/2)^2}}\right. \label{eq15_2} \\
&&\left.+\int_{|{k}/{2}+{2m\Omega}/{k}|}^\infty
   \frac{\beta_kf_{\Omega}^l(p)pdp}{\sqrt{p^2-\left({k}/{2}+{2m\Omega}/{k}\right)^2}}\right]. \nonumber
\end{eqnarray}
%\end{gather}
%
Here we have introduced the coefficients
\begin{gather}\label{eq15_3}\nonumber
\alpha_k=\textmd{Tr}\left(\hat{A}\hat{M}_{\textbf{k}}\hat{A}\hat{M}^\ast_{\textbf{k}}\right)=\left|g_1(M_{\textbf{k}})_{11}u^2+g_2(M_{\textbf{k}})_{22}v^2\right|^2,\\\nonumber
\beta_k=\textmd{Tr}\left(\hat{B}\hat{M}_{\textbf{k}}\hat{A}\hat{M}^\ast_{\textbf{k}}\right)=|uv|^2\left|g_1(M_{\textbf{k}})_{11}-g_2(M_{\textbf{k}})_{22}\right|^2,\\
\gamma_k=\textmd{Tr}\left(\hat{B}\hat{M}_{\textbf{k}}\hat{B}\hat{M}^\ast_{\textbf{k}}\right)=\left|g_1(M_{\textbf{k}})_{11}v^2+g_2(M_{\textbf{k}})_{22}u^2\right|^2,
\end{gather}
and
\begin{eqnarray}%\label{eq15_4}
f^u_{\Omega}(p) &=& f^l_{-\Omega}(p)=n_B(\xi_{\textbf{p}})[1+n_B(\xi_{\textbf{p}})],\nonumber\\
f^u_{-\Omega}(p) &=& n_{\Omega}[n_B(\xi_{\textbf{p}})-n_B(\xi_{\textbf{p}}+2\Omega)]\nonumber \\
&&+n_B(\xi_{\textbf{p}}+2\Omega) [1+n_B(\xi_{\textbf{p}})],\nonumber\\
f^l_{\Omega}(p) &=& \theta(E_{\textbf{p}}-2\Omega)\left\{n_{-\Omega}[n_B(\xi_{\textbf{p}})-n_B(\xi_{\textbf{p}}-2\Omega)]\right. \nonumber \\
&&\left.+n_B(\xi_{\textbf{p}}-2\Omega) [1+n_B(\xi_{\textbf{p}})]\right\}, \label{eq15_4}
%f^u_{\Omega}(p)=f^l_{-\Omega}(p)=n_{\textbf{p}}(1+n_{\textbf{p}}),\\\nonumber
%f^u_{-\Omega}(p)=n_{\Omega}(n_{\textbf{p}}-n^+_{\textbf{p}})+n^+_{\textbf{p}}(1+n_{\textbf{p}}),\\\nonumber
%f^l_{\Omega}(p)=\theta(E_{\textbf{p}}-2\Omega)[n_{-\Omega}(n_{\textbf{p}}-n^-_{\textbf{p}})
%+n^-_{\textbf{p}}(1+n_{\textbf{p}})],
\end{eqnarray}
%where
%\begin{eqnarray}\label{eq15_5}
%n_{\textbf{p}}^{\pm}=\left[\exp\left(E_{\textbf{p}}\pm 2\Omega-\mu\right)/T-1\right]^{-1}.
%\end{eqnarray}

\subsubsection{Bath in the condensed state}

We now consider the bath to be in the Bose condensed phase and obtain the TLS self-energy. The elementary excitations of the
Bose-condensed system are Bogoliubov quasi-particles. An explicit form
of the dispersion law of Bogoliubov excitations depends on the model
used to describe the system of interacting bosons. In the case of small
particle density an appropriate theoretical model is the Bogoluibov
model of weakly-interacting Bose gas. In the framework of this model,
the dispersion law of elementary excitations has the form of
$\epsilon_{\textbf{p}}=\sqrt{E_{\textbf{p}}\left(E_{\textbf{p}}+2g_0n_c\right)}$, where $n_c$ is
particle density in the condensate and $g_0$ the strength of
inter-particle interaction. In the low-energy, long-wavelength limit
$E_{\textbf{p}} \ll 2g_0n_c$ the elementary excitations comprise sound quanta, with
a dispersion $\epsilon_{\textbf{p}}\approx sp$ where $s=\sqrt{g_0n_c/m}$ is the
sound velocity. In the Bose-condensed state, most of the particles are in the condensate but
there are also noncondensate particles, due to both interaction and
finite temperature effects (thermal-excited particles). All
three fractions of particles contribute to
relaxation times of TLS. We consider the quantum limit $T\ll sp$ when
thermal excitations are not important and the theory can be developed
for $T=0$. In the present dilute boson gas limit, the density of the noncondensate particles
is small, and one can neglect the interaction term
due to fluctuations of the condensate density and the noncondensate
density. Thus, the contribution to relaxation times due to the condensate and noncondensate particles can be calculated independently.

\begin{figure} [h]
%\centerline{\input epsf \epsfysize=3.3cm \epsfbox{figure2.pdf}}
\includegraphics[width=8.5cm,angle=0]{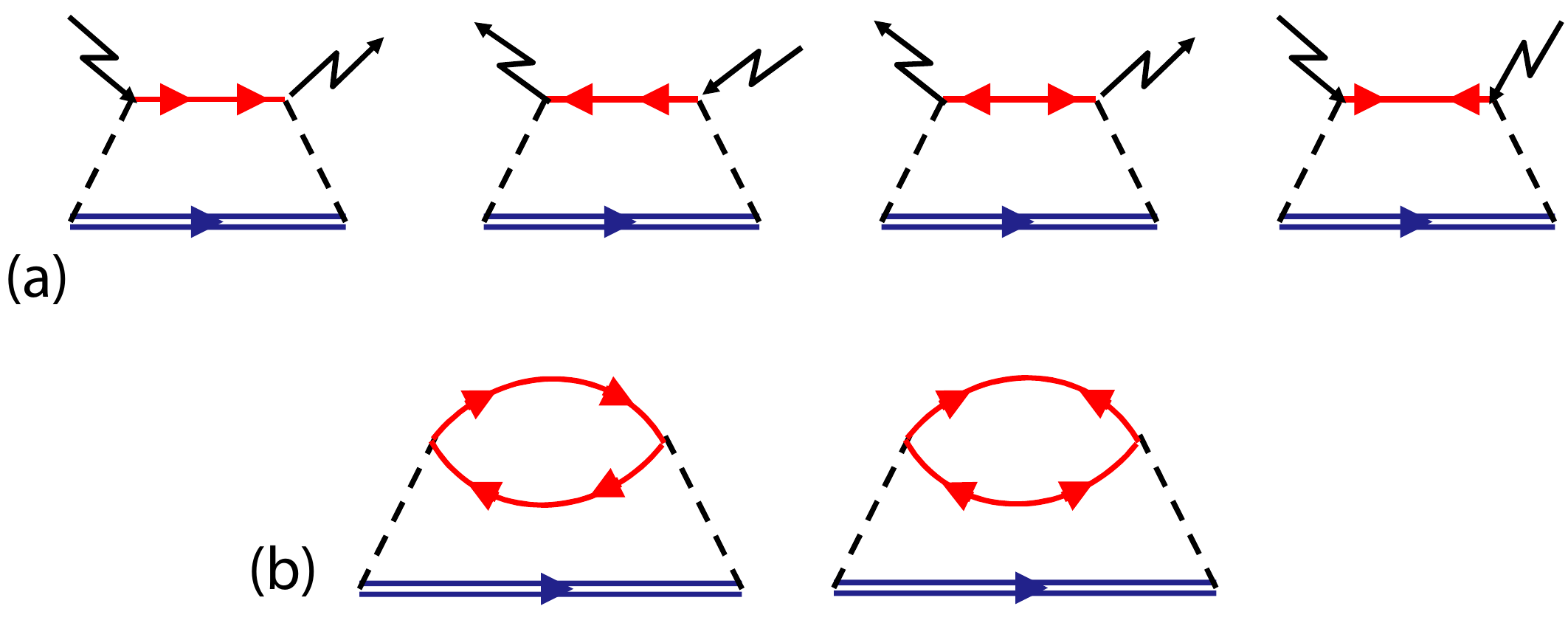}
\caption{Feynman diagrams for TLS self-energy when the bath is
    in the BEC phase. Type (a) describes the condensate particles contribution,
  $\bar{\Sigma}^c$, and (b) is due to the non-condensate particles,
  $\bar{\Sigma}^n$. Double blue line represents the TLS Green
function and dashed lines the TLS electron-bath interaction.  Red lines denote the normal or anomalous Green
  functions of the non-condensate particles, and the zigzag lines 
  stand for $\sqrt{n_c}$ corresponding to the condensate particles.}
%due to the BEC particles and zigzag lines stand for $\sqrt{n_c}$. Other notations like in previous Figure.}
%Feynman diagrams for TLS self-energy in case of BEC
%  bath. Type (a) describes the condensate particles contribution,
%  $\bar{\Sigma}^c$, and (b) is due to the non-condensate particles,
%  $\bar{\Sigma}^n$. Red lines are the normal and anomalous Green
%  functions due to the BEC particles and zigzag lines stand for $\sqrt{n_c}$. Other notations like in previous Figure.}
\label{Figure2}
\end{figure}

The self-energy diagrams in the lowest order with respect to the
TLS-bath coupling are depicted in Fig.~\ref{Figure2}. 
Fig.~\ref{Figure2}(a) corresponds to the contribution to the
  self-energy from the condensate particles and describes virtual processes in which a
condensate particle is scattered by a TLS electron through an
intermediate non-condensate state. Fig.~\ref{Figure2}(b) corresponds
to the contribution from the non-condensate particles and describes polarization of the non-condensate particles induced by the
TLS electrons. These self-energy contributions due to condensate
particles $\hat{\Sigma}^{c R}$ and non-condensate
$\hat{\Sigma}^{nR}$ particles have the following form (see Appendix C)
%The expressions of the self-energies due to condensate $\hat{\Sigma}^{c R}$ and non-condensate
%$\hat{\Sigma}^{nR}$ particles have the following form (see Appendix B)
%
\begin{widetext}
%\begin{gather}\label{eq16}\nonumber
\begin{eqnarray}
\hat{\Sigma}^{c R}(\omega) &=&\frac{n_c}{2ms}\sum_{\textbf{k}}k
\left[\hat{M}_{\textbf{k}}\hat{A}\hat{M}_{-\textbf{k}}\left(\frac{1-n_\Omega}{\omega-\Omega-\epsilon_{\textbf{k}}+i\delta}+\frac{n_\Omega}{\omega-\Omega+\epsilon_{\textbf{k}}+i\delta}\right)
                               \right.\nonumber \\
&&\left.+\hat{M}_{\textbf{k}}\hat{B}\hat{M}_{-\textbf{k}}\left(\frac{1-n_{-\Omega}}{\omega+\Omega-\epsilon_{\textbf{k}}+i\delta}+\frac{n_{-\Omega}}{\omega+\Omega+\epsilon_{\textbf{k}}+i\delta}\right)\right],\label{eq16_1}
\end{eqnarray}
\end{widetext}
\begin{widetext}
\begin{eqnarray}
\hat{\Sigma}^{nR}(\omega) &=&\frac{(ms^2)^2}{2}\sum_{\textbf{k},\textbf{p}}\left[\frac{\hat{M}_{\textbf{k}}\hat{A}\hat{M}_{-\textbf{k}}}{\epsilon_{\textbf{p}}\epsilon_{\textbf{p}+\textbf{k}}}
\left(\frac{1-n_\Omega}{\omega-\Omega-\epsilon_{\textbf{p}}-\epsilon_{\textbf{p}+\textbf{k}}+i\delta}+\frac{n_\Omega}{\omega-\Omega+\epsilon_{\textbf{p}}+\epsilon_{\textbf{p}+\textbf{k}}+i\delta}\right)\right. \nonumber\\
&&\left.+\frac{\hat{M}_{\textbf{k}}\hat{B}\hat{M}_{-\textbf{k}}}{\epsilon_{\textbf{p}}\epsilon_{\textbf{p}+\textbf{k}}}\left(\frac{1-n_{-\Omega}}{\omega+\Omega-\epsilon_{\textbf{p}}-\epsilon_{\textbf{p}+\textbf{k}}+i\delta}+\frac{n_{-\Omega}}{\omega+\Omega+\epsilon_{\textbf{p}}+\epsilon_{\textbf{p}+\textbf{k}}+i\delta}\right)
\right]. \label{eq16_2}
\end{eqnarray}
%\end{gather}
\end{widetext}
Similar steps leading to Eqs.~(\ref{eq15_1})-(\ref{eq15_2}) yields
%The same integrating procedure used above yields
%
%\begin{gather}\label{b10.1}
\begin{eqnarray}%\label{b10.1}
\frac{1}{2\tau^c_u}&=&\frac{1}{2\tau^c_l}
=n_{\Omega}\frac{\pi n_c}{2ms}\sum_kk\beta_k\delta(-2\Omega+\epsilon_{\textbf{k}}),\label{b10.1}\\
\frac{1}{2\tau^n_u}&=&\frac{1}{2\tau^n_l} \label{b10.2} \\
&=& n_{\Omega}\frac{\pi
  (ms^2)^2}{2}\sum_{\textbf{k},\textbf{p}}\frac{\beta_k}{\epsilon_{\textbf{p}}\epsilon_{\textbf{p}+\textbf{k}}}\delta(2\Omega-\epsilon_{\textbf{p}}-\epsilon_{\textbf{p}+\textbf{k}}). \nonumber
\end{eqnarray}
%\end{gather}
%
Explicit expressions for relaxation times depend on the shape of the
wave functions of the TLS upper and lower states through the matrix
elements Eq.~(\ref{eq15}). In the next section we propose and analyze
an experimental setup in which explicit expressions of the relaxation time can be obtained.

\section{Application to coupled QD-dipolar exciton bath} \label{Sec:QD}

Applying our theory developed in the previous sections, we
consider the nanostructure depicted in Fig.~3. A double quantum well
(DQW) 
with closely separated electron-doped and hole doped wells
realizes a 2D gas of indirect excitons (also called dipolar excitons).
A self-assembled QD is positioned above the DQW
and is irradiated with a frequency close to the lowest exciton energy
of the QD.
%so that Rabi oscillations occur between the two levels of QD.
%To describe it we apply the electron-hole representation of QD excitations.
In the electron-hole representation of the QD, the lower quantum state
describes the unexcited QD state (\textit{i.e.} vaccum) $|vac\rangle$,
while the lowest excited QD state describes the state $|eh\rangle$ of an excited electron-hole
pair upon irradiation \cite{QD_TLS1,QD_TLS2,QD_TLS3}. 
%, $|eh\rangle$, is when there is an electron-hole pair in it, excited
%by the external electromagnetic field,
%the frequency of which is close to the electron-hole pair excitation energy.
%energy difference between $|vac\rangle$ and $|eh\rangle$, occurring
%the Rabi oscillations in QD.
%Using an appropriate level in respect of which one can count the energies,
With a shift in energy that does not affect the
system's dynamics, we can assume that these two QD states
correspond exactly to the lower and upper TLS levels in
Eq.~(\ref{eq1}) with $\Delta=(E_e+E_h)/2$, where $E_{e,h}$ are the
%\textbf{By lowest level energies, does it mean the electron and hole energies at the band edge below?} \emph{yes!}
lowest energies of electrons and holes in the QD (corresponding to the
energies at the conduction and valence band edges),
%lowest levels energies of electron and hole in QD,
and $\lambda$ represents the dipole matrix element weighted by the electron-hole
pair envelope wave function integrated over all space.

\begin{figure} [h]
\includegraphics[width=8.5cm,angle=0]{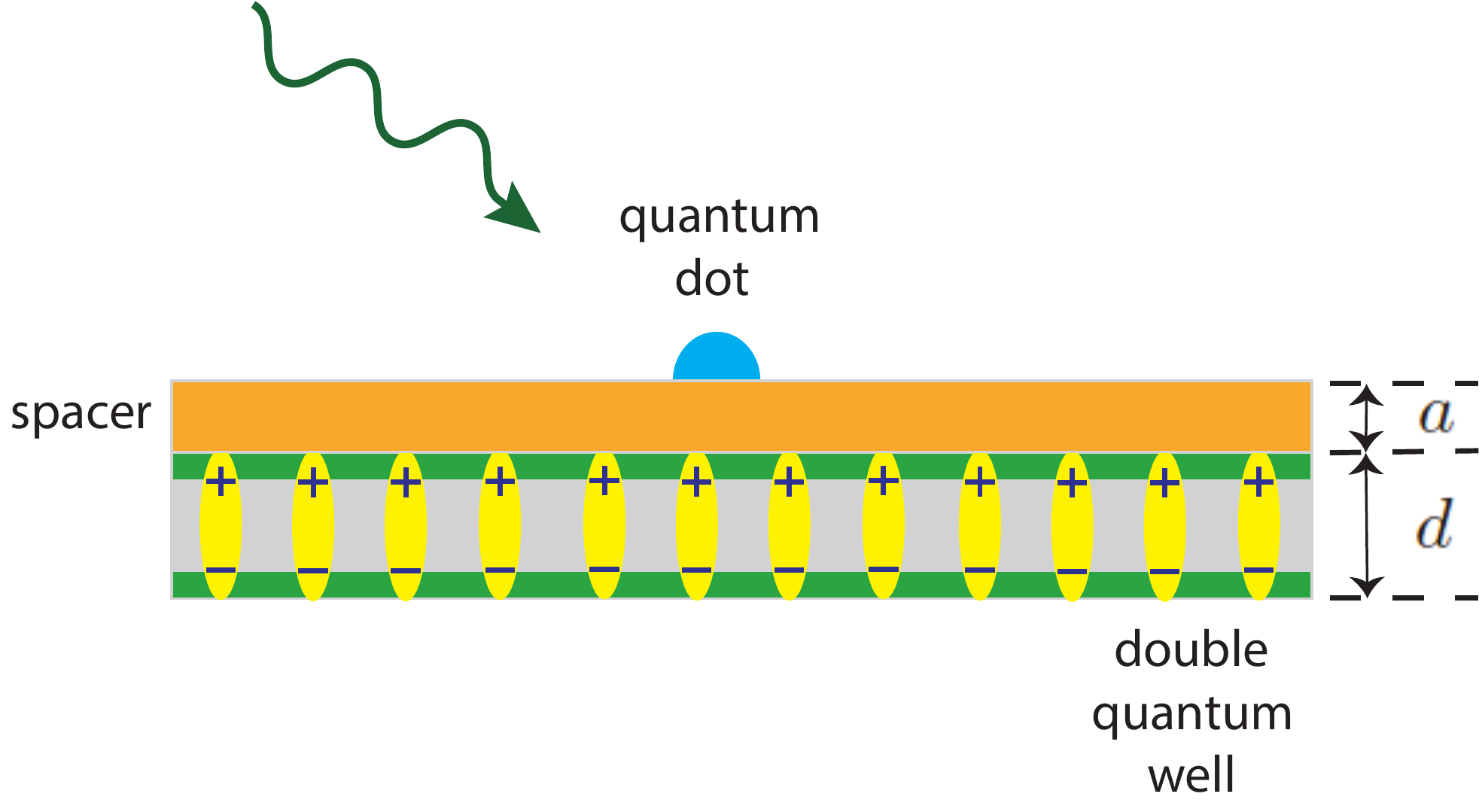}
%\centerline{\input epsf \epsfysize=2cm \epsfbox{figure3.pdf}}
\caption{Schematic of the coupled QD-dipolar excitons system. The quantum dot is
  positioned through a dielectric spacer on top of a double quantum 
  well that hosts the dipolar exciton gas.}
%Sketch of possible experimental setup we consider to be used to detect developed theory.}
\label{Figure3}
\end{figure}

Let us discuss the exciton gas model we use for our calculations. The
indirect excitons
%formed from a pair of electron and hole located in different QWs
have a dipole moment $\textbf{p}$ in the out-of-plane direction of the DQW.
%The electron and hole are located in different QWs interacting via
%Coulomb potential forming the exciton with the dipole moment
%$\textbf{p}$ directed to the normal of the DQW plane.
The excitons are modeled as rigid dipole molecules that are free to
move on the DQW plane as described by the center-of-mass motion of the
dipoles, a valid assumption as long as the dipole's internal degrees of freedom are not excited.
The exciton density $n_{\mathrm{ex}}$ is assumed to be small so that $n_{\mathrm{ex}}a_B^2\ll 1$,
where $a_B$ is an exciton Bohr radius. For simplicity, we
assume that there is no particle tunneling between the QD and the
DQW so that the exchange contribution to the interaction potential is
negligible. This can be guaranteed using a large-band gap dielectric
spacer as a substrate between the QD and the DQW. The direct contribution to the interaction potential between the electron-hole pair in the QD and
the excitons in the DQW is given by
\begin{gather}\label{eq17}
%\label{interactionpotential}\nonumber
U=\int d\textbf{r}\int d\textbf{r}'\left(|\chi_h(\textbf{r})|^2-|\chi_e(\textbf{r})|^2\right)|\varphi(\textbf{r}')|^2u(\textbf{r}-\textbf{r}'),\\
u(\textbf{r}-\textbf{r}')=\frac{e^2}{\varepsilon\sqrt{(\textbf{r}-\textbf{r}')^2+a^2}}-\frac{e^2}{\varepsilon\sqrt{(\textbf{r}-\textbf{r}')^2+(a+d)^2}},
\end{gather}
where $\chi_{e,h}(\textbf{r})$ are the electron (e) and hole (h) wave
functions in the QD, $\varphi(\textbf{r})$ is the QW exciton
center-of-mass wave function, $\varepsilon$ is an effective 
dielectric constant taking account of the dielectric environment of
the spacer and the DQW, $a$ and $d$ are the distance of the QD
to the DQW surface and the separation between the positive charges
and negative charges of the dipole layer (see Fig.~\ref{Figure3}).
%We assume that the QD is sufficiently far away from the
%DQW so that particle tunneling between them is negligible. This implies that the exchange contribution is
%weak and can be neglected as we did in the expression Eq.~(\ref{eq17}).
The Fourier transform of
$u(\textbf{r})$ yields
\begin{gather}\label{eq18}
u(\textbf{k})=\frac{2\pi e^2}{\varepsilon k}\left(1-e^{-kd}\right)e^{-ka}.
\end{gather}
Typical values of the wave vector $k$ here are determined by the
bath excitons of DQW. At temperatures
above the condensation temperature $T>T_c$, the bath in the normal state, the DQW exciton energy is
of the order of $T$ and thus the wave vector $k\sim p_T=\sqrt{2mT}$, 
where $m$ is the exciton mass. At $T=0$,  the bath is in the condensed
phase, the typical value of the bath excitation momentum is of the 
order of $p\leq ms$.
% \textbf{\textit{WK: the following assumption of small
%   separation between the QD and the DQW runs in contradiction with the
%   earlier assumption of large separation that allows tunneling to be
%   neglected, doesn't it? If not, we need to make these two statements
%   more precise so they do not seem to contradict each other. Below, small is compared to $1/p_T$, for the
%   tunneling to be neglected above, what should the separation be
%   compared to?}VK: My idea is like this. Using the form of interaction Eq22 we already assumed that the exchange interaction is negligible, because we used 'a' and 'a+d' instead of particle coordinates across the QWs and integration over it. $kp_T$ is small means that the de Broglie wave of bath's particles is larger then the QD size. I feel that there is no contradiction here, but I don't know how to express this statement correctly. One can just used the contact interaction potential (eq25 below) from the beginning and say 'for simplicity we use the contact interaction between QD and bath'}
Hereafter we assume the distance of the QD
from the DQW as well as the inter-well distance in the DQW to be
sufficiently small so that  $p_Ta, p_Td \ll 1$ and $(ms)a, (ms)d \ll
1$.  This 
%These inequalities
allows us to simplify the expression Eq.~(\ref{eq18}) assuming that 
$kd,ka\ll 1$. Thus we have
\begin{equation}
u(\textbf{k})\equiv U_0=\frac{2\pi e^2d}{\varepsilon},
\,\,\,u(\textbf{r})=U_0\delta(\textbf{r}). \label{eq19}
\end{equation}
Under this approximation the interaction Eq.~(\ref{eq17}) becomes contact-like,
%reduces to 'contact-like' form
%
\begin{equation}
U=U_0\int d\textbf{r}\left(|\chi_h(\textbf{r})|^2-|\chi_e(\textbf{r})|^2\right)|\varphi(\textbf{r})|^2. \label{eq20}
\end{equation} 
In our coupled QD-exciton gas system, since the lower level is the
vacuum,  we have  in Eqs.(\ref{eq8}) and (\ref{eq9}) that the coupling
constants $g_1=U_0,\,\,g_2=0$ and their respective matrix elements $W_{11}=U,\,\,W_{22}=0$.
%Now in Eqs.(\ref{eq8}) and (\ref{eq9}) one can assume that $g_1=u_0,\,\,g_2=0$, and respectively matrix elements are $W_{11}=U,\,\,W_{22}=0$.
%To obtain the analytic expression for Eq.~(\ref{eq20}) we use the parabolic potential confinement to describe the QD. 
The QD is described by a two-dimensional system of electrons and holes
confined by a parabolic potential. For strongly confined QDs where the
QD size $L$ is small compared to the Bohr radius of the electron-hole
pair, the lowest-energy electrons and holes are characterized by the wavefunctions \cite{QD_TLS2} 
%To obtain the analytic expression for Eq.~(\ref{eq20}), we use the
%parabolic potential confinement to describe the QD. 
%the lowest-energy holes
%and electrons are then characterized by the following
%wavefunctions \cite{QD_TLS2} 
$\chi_i(\textbf{r})=\exp(-\rho^2/2a^2_i)/({a_i\sqrt{\pi}})$, where
$a_i$ is an electron $i=e$ or a hole $i=h$ characteristic length 
determined by the QD confinement potential. Using these wavefunctions,
we find the matrix element $(\hat{M}_{\textbf{k}})_{11}=\mathrm{exp}[{-(ka_h/2)^2}]-\mathrm{exp}[{-(ka_e/2)^2}]$.

\subsection{Relaxation times for normal-state bath}
%\emph{Relaxation times for normal-state bath.} %Assuming again that
With $ka_i \ll 1$, we perform the integration over $k$ in
Eqs.~(\ref{eq15_1})-(\ref{eq15_2}) and obtain the following expressions for the
relaxation times
%one can integrate in Eq.~(\ref{eq16}) over $k$.
%Expressions for the relaxation times can be presented in the form
%
%\begin{gather}
\begin{eqnarray}
\frac{1}{2\tau_u}&=&\frac{m\alpha_0}{(2\pi)^2}\int_0^\infty
                     F_+(p,0)n_B(\xi_{\textbf{p}})[1+n_B(\xi_{\textbf{p}})]pdp\nonumber
  \\
&&+\frac{m\beta_0}{(2\pi)^2}\int_0^\infty
   F_+(p,p_0)f^u_{-\Omega}(p)pdp, \label{eq21_1} \\
\frac{1}{2\tau_l}&=&\frac{m\gamma_0}{(2\pi)^2}\int_0^\infty F_-(p,0)n_B(\xi_{\textbf{p}})[1+n_B(\xi_{\textbf{p}})]pdp\nonumber\\
&&+\frac{m\beta_0}{(2\pi)^2}\int_0^\infty
   F_-(p,p_0)f^l_{\Omega}(p)pdp, \label{eq21_2}
%\frac{1}{2\tau_u}&=&\frac{m\alpha_0}{(2\pi)^2}\int_0^\infty F_+(p,0)n_{\textbf{p}}(1+n_{\textbf{p}})pdp+\\\nonumber
%&&+\frac{m\beta_0}{(2\pi)^2}\int_0^\infty
%   F_+(p,p_0)f^u_{-\Omega}(p)pdp, \label{eq21_1} \\
%\frac{1}{2\tau_l}&=&\frac{m\gamma_0}{(2\pi)^2}\int_0^\infty F_-(p,0)n_{\textbf{p}}(1+n_{\textbf{p}})pdp+\\\nonumber
%&&+\frac{m\beta_0}{(2\pi)^2}\int_0^\infty
%   F_-(p,p_0)f^l_{\Omega}(p)pdp, \label{eq21_2}
\end{eqnarray}
%\end{gather}
%
where $F_\pm(p,p_0)=6\pi(p^4\pm p^2p_0^2+p_0^4/6)\theta(p^2\pm
p_0^2)$ with $p_0^2=4m\Omega$, and
%where $$F_\pm(p,p_0)=6\pi[p^4\pm p^2p_0^2+p_0^4/6]\theta[p^2\pm
%p_0^2],\,\,\,p_0^2=4M\Omega$$ and
\begin{eqnarray}
\alpha_0 &=& U_0^2(a_h^2-a_e^2)^2u^4/16,\nonumber \\
\beta_0 &=& U_0^2(a_h^2-a_e^2)^2|uv|^2/16,\nonumber \\
\gamma_0 &=& U_0^2(a_h^2-a_e^2)^2v^4/16. \label{eq21_3}
\end{eqnarray}
To
simplify Eqs.~(\ref{eq21_1})-(\ref{eq21_2}) we consider two limiting cases. 
%\textbf{WK: I think the `high temperature limit' is
%  contradictory to the regime we have started out considering, which
%  is always low temperature. We said before: 'We consider the quantum
%  limit $T \ll sp$ when
%thermal excitations are not important and the theory can
%be developed for T = 0.' I'd suggest, should we remove the following high
%  temperature result?}\texttt{We have normal and BEC regimes. The
%  values of $\Omega$ and $T_c$ can be in principle, arbitrary. Thus,
%  In normal state one can consider two limits $T_c<T<<\Omega$ or
%  $T_c<\Omega<<T$. In BEC regime $T<T_c$ but for simplicity, we do
%  not consider the thermal-activated Bogliubov excitations. To
%  disregard it one need to have $T<<sp,\,\,T<T_c$ independent on
%  relation between $\Omega$ and $T$.} 
%For weak TLS-light coupling such that $\Omega\ll T$ [with a
%temperature still small compared to $\hbar^2/(2m d^2)$], 
At large temperatures $T \gg\Omega$ [while still small compared to
$\hbar^2/(2m d^2)$],  
we can keep up to the zeroth order in $\Omega/T$ 
%For weak TLS-light coupling such that $\Omega\ll T$, 
%one can put $\Omega\rightarrow 0$ in the distribution functions
%Eq.~(\ref{eq15_4}). Thus $p_0^2=0$, 
with $F_{\pm}(p,p_0) \approx 6\pi p^4$ and 
%$n_{\pm\Omega}$ drops out, and 
$f^u_{-\Omega}(p)\approx f^l_{\Omega}(p)\approx
n_B(\xi_{\textbf{p}})[1+n_B(\xi_{\textbf{p}})]$ in
Eqs.~(\ref{eq21_1})-(\ref{eq21_2}). The relaxation rates for
$\Omega\ll T$ 
%at high temperatures 
then read
\begin{equation}
\frac{1}{\tau_{u,l}} 
=\frac{1}{\tau_T}\left(1\pm \frac{\varepsilon_0}{\Omega}\right)\mathrm{Li}_2\left(e^{\mu/T}\right),
%\int_0^\infty\frac{xdx}{e^{x-(\mu/T)}-1}, 
\label{eq22}
\end{equation}
%\begin{gather}
% \begin{eqnarray}
% \left(
%   \begin{array}{c}
%     \tau^{-1}_u \\
%     \tau^{-1}_l \\
%   \end{array}
% \right)
% &=&\frac{1}{\tau_T}\left(
%   \begin{array}{c}
%     u^2 \\
%     v^2 \\
%   \end{array}
% \right)\mathrm{Li}_2\left(e^{\mu/T}\right),
% %\int_0^\infty\frac{xdx}{e^{x-(\mu/T)}-1}, 
% \label{eq22}
% \end{eqnarray}
%\end{gather}
%
where the $+,-$ signs apply for $\tau_{u,l}$ respectively, and 
\begin{equation}
\frac{1}{\tau_T} =
\frac{3 U_0^2 m^4T^3(a_h^2-a_e^2)^2}{4\pi\hbar^9}, \label{eq22_1}
\end{equation}
and $\mathrm{Li}_2(x)$ is the polylogarithm function of order $2$. 
In the opposite limit of low temperatures $T \ll \Omega$,  
%In the opposite limit of strong TLS-light coupling $\Omega\gg T$, 
we keep up to the zeroth order in $T/\Omega$. Then the respective first
terms in Eqs.~(\ref{eq21_1})-(\ref{eq21_2}) drop out, and we have 
%we have
$f^u_{-\Omega}(p)\approx n_\Omega n_B(\xi_{\textbf{p}})$ and
$f^l_{\Omega}(p)\approx n_\Omega \theta(E_{\textbf{p}}-2\Omega)
n_B(\xi_{\textbf{p}}-2\Omega)$ in the second terms. %In this limit the second integrals in Eqs.~(\ref{eq21_1})-(\ref{eq21_2}) are easilycalculated. 
The result is 
\begin{equation}
    \frac{1}{\tau_{u}} = \frac{1}{\tau_{l}} 
=\frac{1}{\tau_T}n_{\Omega}\frac{\pi^2}{3}\left(\frac{\lambda}{T}\right)^2\frac{\hbar^2
    n_{\mathrm{ex}}}{mT},  \label{eq23}
\end{equation}
%\begin{gather}
% \begin{eqnarray}\label{eq23}
% \left(
%   \begin{array}{c}
%     \tau^{-1}_u \\
%     \tau^{-1}_l \\
%   \end{array}
% \right)
% &=&\frac{1}{\tau_T}n_{\Omega}|uv|^2\frac{2\pi^2}{3}\left(\frac{\Omega}{T}\right)^2\frac{\hbar^2
%     n_{\mathrm{ex}}}{mT},  
% %&&\left.+
% %\left(
% %  \begin{array}{c}
% %    u^4 \\
% %    v^4 \\
% %  \end{array}
% %\right) \mathrm{Li}_2\left(e^{\mu/T}\right)\right],
% %\int_0^\infty\frac{xdx}{e^{x-(\mu/T)}-1}\right],
% \end{eqnarray}
%\end{gather}
where $n_{\mathrm{ex}}$ is the exciton density in the bath. 
%The first term here gives the main contribution because of the large
%factor $\Omega/T\gg1$. 
We have restored $\hbar$ in the expressions for the relaxation rates here and in
the following section. 

\subsection{Relaxation times for Bose-condensed bath}
%\emph{Relaxation times for Bose-condensed bath.}
Straightforward but
cumbersome integration in Eqs.~(\ref{b10.1})-(\ref{b10.2}) results in the following
expressions for relaxation times due to the condensate and non-condensate particles in the bath
\begin{equation}
%\frac{1}{2\tau^c_u}=\frac{1}{2\tau^c_l}
%=n_{\Omega}\frac{|\lambda|^2n_cu_0^2}{4Ms^4}M^2_{2\Omega/s}, \label{eq24}
\frac{1}{2\tau^c_u}=\frac{1}{2\tau^c_l}
=n_{\Omega}\frac{|\lambda|^2n_cU_0^2}{4\hbar^3 ms^4}\left[e^{-(\Omega a_h/s)^2}-e^{-(\Omega
    a_e/s)^2}\right]^2, \label{eq24}
\end{equation}
%
%where $M_{2\Omega/s}=\left(e^{-(\Omega a_h/s)^2}-e^{-(\Omega
%    a_e/s)^2}\right)^2$
and
\begin{eqnarray}
&&\frac{1}{2\tau^n_u} =\frac{1}{2\tau^n_l}
=n_{\Omega}\frac{\vert\lambda\vert^2 m^2sU_0^2}{16\pi^2\hbar^4 \Omega^2}\label{eq25} \\
%=n_{\Omega}|uv|^2\frac{sm^2U_0^2}{4\pi^2}\label{eq25} \\
%=n_{\Omega}|uv|^2\frac{2(\pi Ms^2u_0)^2}{(2\pi s)^3}\times\\\nonumber
&&\times\left[\frac{F(\sqrt{2}\Omega
   a_h/s)}{a_h\sqrt{2}}+\frac{F(\sqrt{2}\Omega
   a_e/s)}{a_e\sqrt{2}}-2\frac{F(\sqrt{a_h^2+a_e^2}\Omega/s)}{\sqrt{a_h^2+a_e^2}}\right], \nonumber
\end{eqnarray}
where $F(x)$ is the Dawson integral
$$
F(x)=e^{-x^2}\int_0^x e^{t^2}dt.
$$

\section{Discussion}
Simple analysis of Rabi oscillations given above accounting for the finite values of relaxation times yields
\begin{eqnarray}
|\langle\psi_2^+(0)\psi_1(t)\rangle|^2 &=&\frac{|\lambda|^2}{2\Omega^2}
\left(1-\cos\Omega
  t\right)e^{-{t}\left({1}/{2\tau_u}+{1}/{2\tau_l}\right)}\nonumber \\
&&+\frac{|\lambda|^2}{4\Omega^2}\left(e^{-{t}/{2\tau_u}}-e^{-{t}/{2\tau_l}}\right)^2. \label{eq26}
\end{eqnarray}

We focus our discussion on the low temperature regime $T \ll \Omega$. 
First we note that the relaxation rates for the upper and lower levels
coincide both in the normal phase [Eq.~(\ref{eq23})] and in the
Bose-condensed phase  [Eqs.~(\ref{eq24})-(\ref{eq25})]. With relaxation times for the
upper and lower levels being equal, Eq.~(\ref{eq26}) is simplified to 
\begin{gather}\label{eq28}\nonumber
|\langle\psi_2^+(0)\psi_1(t)\rangle|^2=\frac{|\lambda|^2}{2\Omega^2}\left(1-\cos\Omega t\right)e^{-t/\tau},
\end{gather}
where $1/\tau=1/\tau_u=1/\tau_l$, with $1/\tau$ given by
Eq.~(\ref{eq23}) in the normal phase and by $1/\tau^c+1/\tau^n$ from
Eqs.~(\ref{eq24})-(\ref{eq25}) in the BEC phase. Secondly, 
the relaxation rates are all proportional to the distribution function $n_{\Omega}$ of the dressed quasiparticle
states given in Eq.~(\ref{nOmega}). Since $n_{\Omega} =
1$ only if the upper dressed quasiparticle state $+\Omega$ is
occupied and vanishes otherwise, finite relaxation of the TLS Rabi
oscillations occurs only when the pump field frequency exceeds the TLS
energy level difference $\omega \geq 2\Delta$ \cite{Remark1}.  
In our following discussion, therefore, we focus on the regime $\omega
\geq 2\Delta$.  
%and compare the relaxation rates in the normal phase and the BEC phase. 
In the normal phase, we note that the low-temperature relaxation rates
Eq.~(\ref{eq23}) are independent of the driving frequency and
increases monotonically with the TLS-light coupling as $\lambda^2$. In the BEC
phase, we find that the relaxation rates Eqs.~(\ref{eq24})-(\ref{eq25}) exhibit 
strong non-monotonic dependence on the driving frequency through the Rabi
frequency and the TLS-light coupling. 

\begin{figure} [h]
\includegraphics[width=8.5cm,angle=0]{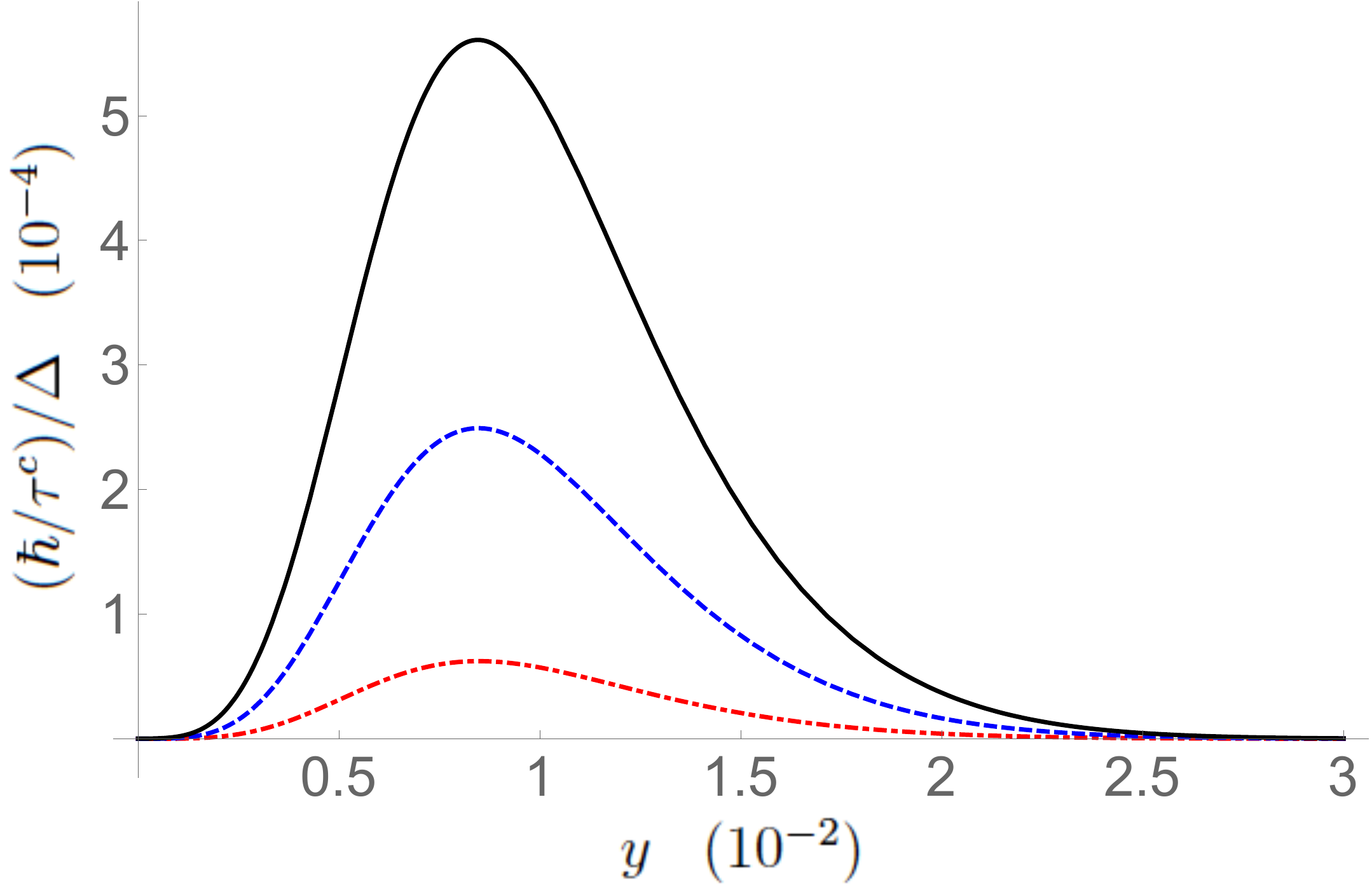}
%\centerline{\input epsf \epsfysize=5cm \epsfbox{figure4.pdf}}
\caption{Relaxation rate $\hbar/\tau^{c}$ due to condensate particles (normalized by
  $\Delta$) versus frequency detuning $y$. Red (dot-dashed), blue
  (dashed) and black (solid) lines correspond to $\lambda=0.1\,\mathrm{meV}$, $0.2\,\mathrm{meV}$ and $0.3\,\mathrm{meV}$ respectively.}
\label{Figure4}
\end{figure}
\begin{figure} [h]
\includegraphics[width=8.5cm,angle=0]{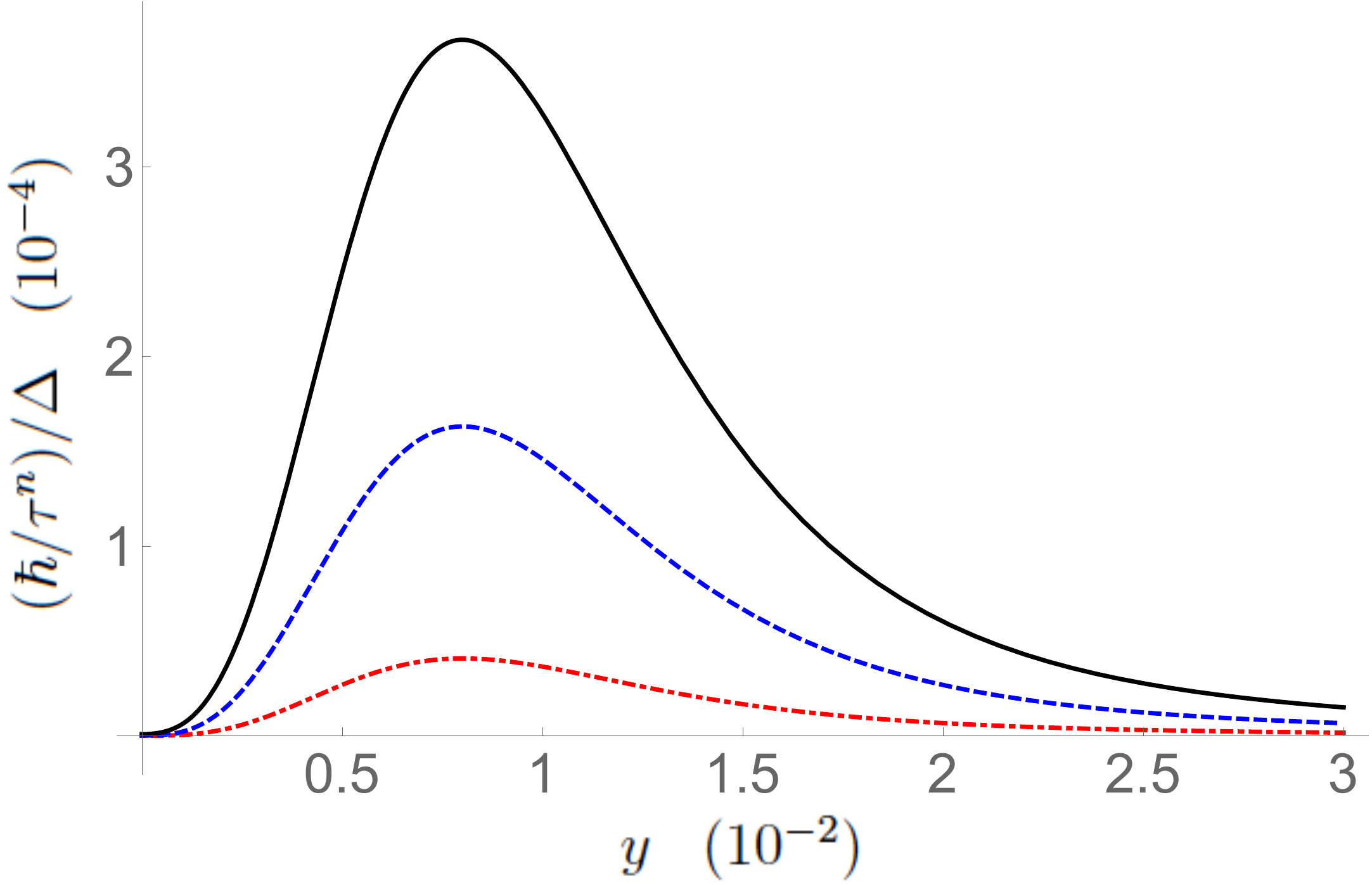}
%\centerline{\input epsf \epsfysize=5cm \epsfbox{figure5.pdf}}
\caption{Relaxation rate $\hbar/\tau^{n}$ due to non-condensate particles (normalized by
  $\Delta$) 
  versus frequency detuning $y$. Red (dot-dashed), blue
  (dashed) and black (solid) 
  lines correspond to $\lambda=0.1\,\mathrm{meV}$, $0.2\,\mathrm{meV}$ and $0.3\,\mathrm{meV}$ respectively.}
\label{Figure5}
\end{figure}
\begin{figure} [h]
\includegraphics[width=8.5cm,angle=0]{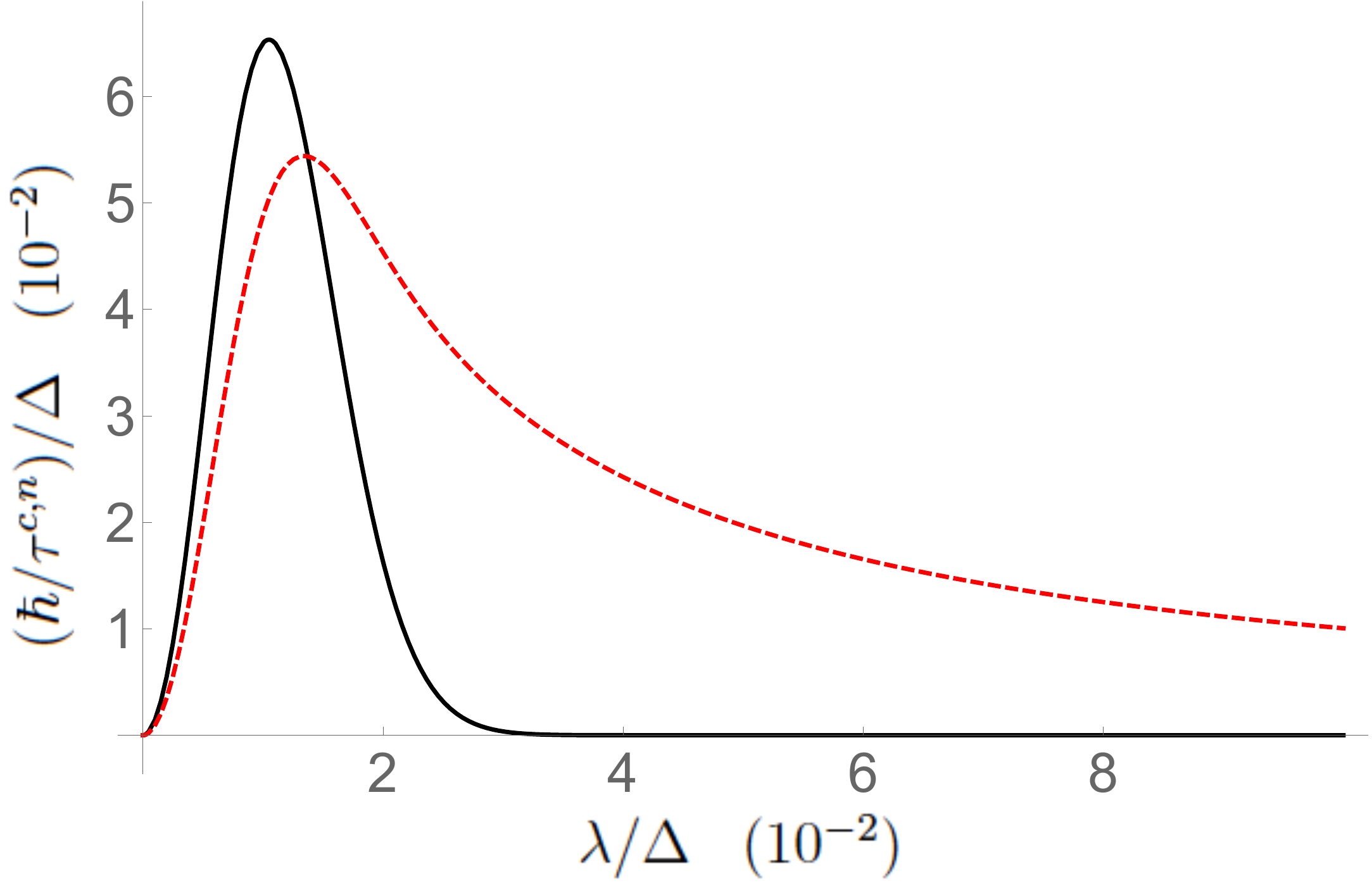}
%\centerline{\input epsf \epsfysize=5cm \epsfbox{figure5.pdf}}
\caption{Relaxation rates due to condensate and non-condensate
  particles versus TLS-light coupling $\lambda/\Delta$ at frequency detuning
  $y = 0.01$. Solid (black) line corresponds to $\hbar/\tau^{c}$ and
  dashed (red) line to $\hbar/\tau^{n}$, respectively.}
\label{Figure6}
\end{figure}
\begin{figure} [h]
\includegraphics[width=8.5cm,angle=0]{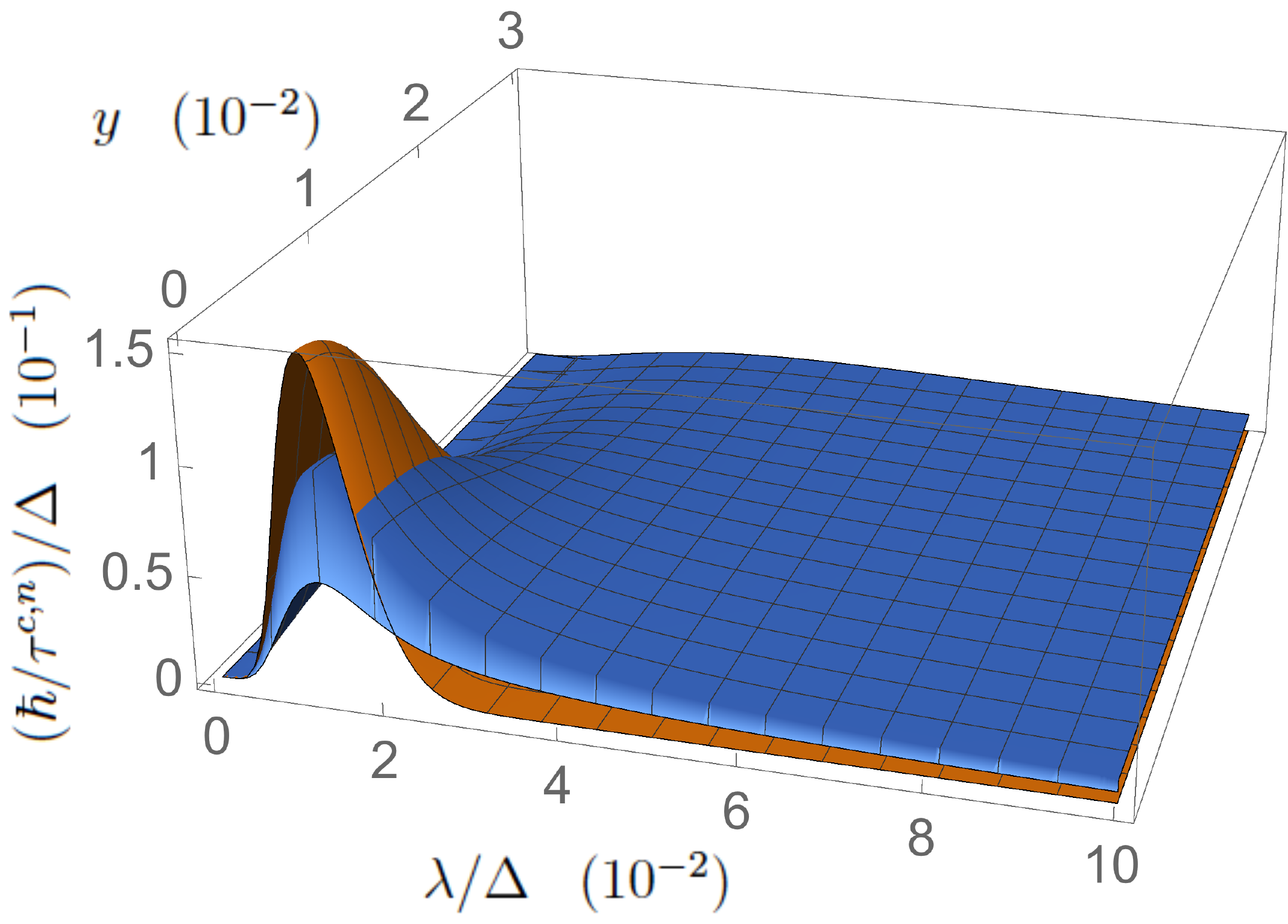}
%\centerline{\input epsf \epsfysize=5cm \epsfbox{figure5.pdf}}
\caption{Three-dimensional plot of relaxation rates $\hbar/\tau^{c}$
  and $\hbar/\tau^{n}$ as a function of TLS-light coupling $\lambda/\Delta$
  and frequency detuning $y$. Grey (red) surface corresponds to
  $\hbar/\tau^{c}$ and black (blue) surface to  $\hbar/\tau^{n}$, respectively.}
\label{Figure7}
\end{figure}

Below, we proceed to analyze the numerical dependence of the relaxation rates on the driving
frequency and TLS-light coupling in the BEC phase. 
For the TLS, we take the following parameters
  $\Delta = 500 \,\mathrm{meV}$, $m_e=0.067\,m_0$ and $m_h=0.45\,m_0$ ($m_0$ is the electron mass) typical for GaAs-based QDs \cite{GaAs_QD1,GaAs_QD2,GaAs_QD3}. The characteristic lengths of the hole and the electron wavefunctions are taken as $a_h = 2\,
\mathrm{nm}$ and $a_e \approx a_h \sqrt{m_h/m_e} = 2.23 a_h$. With the dipole matrix element of the QD $\sim
10\,-\,100\,\mathrm{Debye}$ ($1\,\mathrm{Debye} = 3.3\times
10^{-30}\,\mathrm{Cm}$) and optical field strength $0.1\,-\,10\,\mathrm{MVm^{-1}}$, the TLS-light coupling constant takes
the range of values $\lambda \sim 0.1\,-\,10\,\mathrm{meV}$. 
For the dipolar exciton gas, we take $d=10\,\mathrm{nm}$, $n_c=10^{10}\,\mathrm{cm}^{-2}$, $\varepsilon = 12.5$ and
$m = m_e+m_h = 0.517\,m_0$ typical for GaAs DQW structures. 
%(we take the effective dielectric constant to be that  of GaAs). 
%Assuming a simple point-charge treatment of dipolar excitons, we use 
%$g(r) = (2e^2/\varepsilon)(1/r-1/\sqrt{r^2+d^2})$ \cite{Zimmermann} as the form of
%the exciton-exciton interaction for estimating $g_0$. 
To provide an estimate for the inter-particle interaction $g_0$, we assume a simple point-charge
treatment of the dipolar excitons, and the exciton-exciton interaction
potential takes the form \cite{Zimmermann} $g(r) =
(2e^2/\varepsilon)(1/r-1/\sqrt{r^2+d^2})$. 
Fourier transform of $g(r)$ then gives $g(k) = (4\pi e^2/\varepsilon k)[1-\mathrm{exp}(-kd)]$ and the
coupling constant $g_0 \equiv g(k = 0) = 4\pi e^2 d/\varepsilon$. The
Bogoliubov speed of sound $s$ is thus also fixed from $s = \sqrt{g_0
  n_c/m}$. 

For convenience, we display the frequency $\omega$ in terms of the
dimensionless frequency detuning
$y=(\omega-2\Delta)/(2\Delta)$. Figs.~\ref{Figure4}-\ref{Figure5} show
the relaxation rates $\hbar/\tau^{c}$ and $\hbar/\tau^{n}$ as a
function of $y$ for relatively small values of $\lambda \sim 0.1\,\mathrm{meV}$. 
%
% In Figs.~\ref{Figure4}-\ref{Figure5}, we plot the relaxation rates $1/\tau^{c}$
% and $1/\tau^{n}$ as a function of $y$ for different values of $\lambda$. It is convenient to introduce dimensionless
% quantities. We use $y=(\omega-2\Delta)/(2\Delta)$ as the
% dimensionless frequency detuning to express the Rabi frequency as
% $\Omega=\sqrt{(\Delta y)^2+\lambda^2}$ and express the relaxation rates
% in units of $\Delta$: $\hbar /(\tau^c\Delta)$ and
% $\hbar/(\tau^n\Delta)$. 
%
We find that both relaxation rates behave non-monotonically as a function of detuning,
reaching maximum values at $y \sim 0.01$ and then becoming exponentially suppressed
at larger values of $y$. $\hbar/\tau^{c}$ is more strongly suppressed than
$\hbar/\tau^{n}$. Secondly, we observe that, for the present values of $\lambda \sim 0.1\,\mathrm{meV}$, the condensate and non-condensate fractions contribute to
the relaxation rate by the same order of magnitude, with
$\hbar/\tau^{c}$ exceeding $\hbar/\tau^{n}$. This trend is maintained
until $\lambda$ reaches $\sim 1\%$ of $\Delta$ (corresponding to $5\,\mathrm{meV}$), when $\hbar/\tau^{c}$
starts to drop signifcantly faster than 
$\hbar/\tau^{n}$. Fig.~\ref{Figure6} shows both quantities plotted
versus $\lambda/\Delta$ at a frequency detuning $y = 0.01$, from which we
observe that $\hbar/\tau^{c}$ decreases much more abruptly than
$\hbar/\tau^{n}$. When $\lambda$ is increased beyond $\sim 0.02\Delta$, it is seen that
$\hbar/\tau^{n}$ now overtakes $\hbar/\tau^{c}$. For $\lambda$ values
beyond $0.03\Delta$, $\hbar/\tau^{c}$ has dropped essentially to
zero and the non-condensate fraction constitutes the dominant
contribution to relaxation. 

Although the relaxation rates vanish
expectedly %\cite{Remark2} 
when $\lambda = 0$, they do not vanish at
zero frequency detuning $y = 0$, as one might conclude by inspecting
Figs.~\ref{Figure4}-\ref{Figure5}. To examine more fully the behavior of
$\hbar/\tau^{c}$ and $\hbar/\tau^{n}$, we plot them in Fig.~\ref{Figure7} in the full range of $\lambda$
and $y$. At $y = 0$, both relaxation rates become small only when
$\lambda \ll \Delta$; in addition $\hbar/\tau^{c}$ also 
become small when $\lambda \gtrsim 0.03\Delta$. Around $\lambda \sim
0.01\Delta$, both relaxation rates as a function of $y$ reach maximum at $y = 0$.   

%Interestingly, we also find that the relaxation rates
%$1/\tau^{c}, 1/\tau^{n}$ exhibit a non-monotonic dependence on the
%TLS-light coupling $\lambda$, increases at first and then
%decreases. At a fixed frequency therefore, one can tune the relaxation 
%rates by changing the amplitude of the optical pump field. 

% strongly non-monotonic functions of external
% driving field frequency. Importantly, we note that the main contribution to the Rabi oscillations damping comes from the
% non-condensate particles being $\sim10^3$ times larger
% (Fig.~\ref{Figure5}) than the condensate particles
% (Fig.~\ref{Figure4}). 

Conventionally, the BEC phase transition of a dipolar exciton gas is
detected using optical spectroscopy. In the BEC phase, the excitons or exciton-polaritons are described by a
single coherent wave function and emit light coherently. The resulting luminescence peak becomes much narrower in 
comparison with that in the normal phase, signaling formation of the condensate state. Another way to detect the BEC phase transition has been theoretically
suggested recently~\cite{10_RefKovalevChaplikJETP2016,10_1RefKovalevChaplikJETPLetters2016}. BEC phase transition strongly influences the non-equilibrium
properties of a dipolar exciton gas driven by an external surface acoustic waves (SAW). Under phase transition, the SAW attenuation
effect and the SAW-exciton drag current become strongly modified,
allowing one to detect the BEC phase transition using acoustic
spectroscopy. On top of the foregoing, our findings in principle provide a new strategy to detect the BEC
phase transition of the dipolar exciton gas. While the QD's relaxation rate 
displays only a monotonic linear dependence on the light 
intensity when the exciton gas is in the normal phase, 
it becomes strongly non-monotonic as a function of both the pump
field's frequency and intensity once the exciton gas is in the BEC phase. Thus, by monitoring
the Rabi oscillation dynamics of the QD, the normal and condensed phases of the
exciton gas can be distinguished by the dependence of the relaxation rate on the
frequency and intensity of the driving field.

% The idea is that BEC phase transition strongly influences the non-equilibrium
% properties of a dipolar exciton gas driven by an external surface
% acoustic waves (SAW). The SAW attenuation effect and SAW-exciton drag
% effect have been considered both in the normal and the BEC phase of
% exciton dipolar gas. Strong modifications of the SAW attenuation
% coefficient and the SAW-exciton drag current as a function of the
% exciton density and SAW frequency under phase transition allows one to
% experimentally detect the BEC phase transition using acoustic spectroscopy. 

\section{Conclusion}
To conclude, we have developed a theory for the relaxation of optically pumped
two-level systems coupled to a bosonic bath using the nonequilibrium 
Keldysh technique and the diagrammatic perturbation theory. To elucidate the effects of bath phase
transition, we have considered the cases when the bosonic bath is in the
normal state and in the Bose-condensed state. 
%It was shown that when the bath is in the BEC state, Rabi 
%oscillations is relaxed mainly via interaction with non-condensate bath
%particles. 
We then apply our theory to study the scenario of an illuminated quantum
dot coupled to a dipolar exciton gas. The condensate and non-condensate fractions of the bath particles 
contribute to the relaxation rate by variable proportions depending on
the value of pump field amplitude. When the
pump field is weak, both fractions contribute by about the same order
of magnitude; while for strong pump field, the non-condensate fraction
becomes the dominant contribution. Our findings also show that the phase transition of the
dipolar exciton gas to the BEC regime results in a strong dependence of the relaxation
rate on the optical pump field. %While the relaxation rate in the normal phase is 
The relaxation rate then exhibits a strong non-monotonic behavior, reaching a maximum and then becoming exponentially
suppressed as a function of both the pump field's frequency and amplitude.  
%The relaxation rate is exponentially
%suppressed by the driving frequency and exhibits a non-monotonic
%dependence as a function of both the pump field's frequency and field strength. 
%non-monotonic dependence of the 
%relaxation rate as a function of the frequency of the 
%driving electromagnetic field. 
Such a non-monotonic dependence could in principle serve as a smoking
gun for detecting BEC phase transition of the coupled dipolar exciton
gas. Finally, we point out that despite our focus on dipolar exciton
gas in this work, the theory we have developed is also applicable to 
other types of Bose gas, such as 2D exciton-polaritons~\cite{10_7Deng2010}, magnons~\cite{10_8Pokrovskii2013} and cold
atoms~\cite{10_9Pethick2002}. 

\section{Acknowledgments}
V.M.K. acknowledges the support from RFBR grant $\#16-02-00565a$.
W.K. acknowledges the support by a startup fund from the University of
Alabama.

\section{Appendix} \label{Appendix}

\subsection{Non-Equilibrium Green's Functions} \label{Sec:NEGF}

Because of the time-dependent perturbation from light, we employ the
Keldysh formalism to calculate the Green's function and distribution function of
the system. Following established routes in non-equilibrium
Green's function formalism, the left-multiplied and
right-mulitplied Dyson equations for the contour-ordered Green's
function ${G}^c$ are
\begin{eqnarray}
G_0^{-1} {G}^{c} &=& 1+{\Sigma}^{c} {G}^{c}, \\
{G}^{c} G_0^{-1} &=& 1+{G}^{c} {\Sigma}^{c}. \label{Dyson}
\end{eqnarray}
We are interested in the Green's function of the TLS under
irradiation, therefore the self-energy
${\Sigma}$ due to interaction with the bath is set to zero. In
the rotating frame, we already find
\begin{gather}\label{5}
\hat{G}_0^{-1}(t,t') = \left(
  \begin{array}{cc}
    i\partial_t-\left(\Delta-\frac{\omega}{2}\right) & -\lambda  \\
    -\lambda^\ast  & i\partial_t+\left(\Delta-\frac{\omega}{2}\right) \\
  \end{array}
\right).
\end{gather}
First we derive the retarded Green's function. Applying Langreth's
rules \cite{14_Jauhobook} to the two equations in Eq.~(\ref{Dyson}) and summing them
together, we have $\hat{G}_0^{-1}\hat{G}^{R}+\hat{G}^{R}\hat{G}_0^{-1}
= 2\delta(t-t')$,
\begin{eqnarray}
%\bar{G}_0^{-1}\bar{G}^{R}+\bar{G}^{R}\bar{G}_0^{-1} &=& 2\delta(t-t')
%\nonumber
%\\
\left(i\frac{\partial \hat{G}^R}{\partial t}-\hat{H}_0
  \hat{G}^R\right)+\left(-i\frac{\partial \hat{G}^R}{\partial
    t'}-\hat{G}^R \hat{H}_0\right) =  2\delta(t-t'). \nonumber \\
\label{Dyson2}
\end{eqnarray}
We transform the time variables $t,t'$, into the Wigner coordinates with the average time $T = (t+t')/2$ and relative time $\tau =
t-t'$. %Noting that ${\partial}/{\partial t} = ({1}/{2}){\partial}/{\partial
%  T}+{\partial}/{\partial \tau}$ and ${\partial}/{\partial t'} = ({1}/{2}){\partial}/{\partial
%  T}-{\partial}/{\partial \tau}$,
%This is the place where it needs to be more careful and was missed before: we need to separate the slowly varying and fast
%varying time variables before Fourier transformation. Fourier
%transformation is done only w.r.t. the fast time variable $\tau =
%t-t'$, leaving the slow time variable $T = (t+t')/2$ untouched.
% Since
% %
% \begin{eqnarray}
% \frac{\partial}{\partial t} = \frac{1}{2}\frac{\partial}{\partial
%   T}+\frac{\partial}{\partial \tau}, \\
% \frac{\partial}{\partial t'} = \frac{1}{2}\frac{\partial}{\partial
%   T}-\frac{\partial}{\partial \tau}
% \end{eqnarray}
% %
Eq.~(\ref{Dyson2}) becomes
\begin{eqnarray}
i2\frac{\partial \hat{G}^R}{\partial
  \tau}-\left\{\hat{H}_0,\hat{G}^R\right\} = 2\delta(\tau).
\end{eqnarray}
Performing Fourier transformation with respect to
$\tau$
%Now Fourier transforming $\int\,\mathrm{d}\tau\,exp(-i \varepsilon\tau)$
gives
\begin{eqnarray}
2\varepsilon \hat{G}^R-\left\{\hat{H}_0,\hat{G}^R\right\} = 2.
\end{eqnarray}
Solving this matrix equation yields the retarded Green's function
 in Eq.~(\ref{eq5}):
\begin{eqnarray}
\hat{G}^R(\varepsilon,T) &=&\frac{1}{2\Omega}\left(
                                          \begin{array}{cc}
                                            \varepsilon+\varepsilon_0 & \lambda \\
                                            \lambda^\ast & \varepsilon-\varepsilon_0 \\
                                          \end{array}
                                        \right) \nonumber \\
&&\times
  \left[\frac{1}{\varepsilon-\Omega+i\delta}-\frac{1}{\varepsilon+\Omega+i\delta}\right]
  \nonumber \\
&=&\hat{A}\frac{1}{\varepsilon-\Omega+i\delta}+\hat{B}\frac{1}{\varepsilon+\Omega+i\delta}, \label{7}
\end{eqnarray}
with $\hat{A},\hat{B}$ defined in Eq.~(\ref{eq5_1}).

Now from the contour-ordered Dyson's equations Eq.~(\ref{Dyson}) and
applying Langreth's rule, then subtracting the two equations, we get
$G_0^{-1}G^{<}-G^{<} G_0^{-1} = 0$,
\begin{eqnarray}
\left(i\frac{\partial \hat{G}^{<}}{\partial
    t}-\hat{H}_0 \hat{G}^{<}\right)-\left(-i\frac{\partial \hat{G}^{<}}{\partial
    t'}-\hat{G}^{<}\hat{H}_0\right) = 0,
%\bar{G}_0^{-1}\bar{G}^{<} = i\frac{\partial \bar{G}^{<}}{\partial
%    t}-\bar{H}_0 \bar{G}^{<}, \\
%\bar{G}^{<} \bar{G}_0^{-1}= -i\frac{\partial \bar{G}^{<}}{\partial
%    t'}-\bar{G}^{<}\bar{H}_0,
\end{eqnarray}
Transforming into the Wigner coordinates, we obtain the kinetic equation
\begin{equation}
i\frac{\partial \hat{G}^{<}}{\partial t}-\left[\hat{H}_0,\hat{G}^{<}\right]
= 0
\end{equation}
The density matrix $f(t)$ is given by the equal-time Keldysh Green's function
$\hat{f}(t) = -i\hat{G}^{<}(t,t) = -i\hat{G}^{<}(T = t, \tau = 0)$, which satisfies
% Note that, it may seem fishy to do the unitary transformation first to
% the rotating frame, and then transform into the variable $\tau$ and
% $T$. Afterall, the unitary transformation only depends on $t$, and we use either $t,t'$ or $T,\tau$ as two independent
% sets of variables, but not mixing them together. But this is ok, since
% in the end we set $t=t'$ in the kinetic equation as we are
% interested in the density matrix. Setting $t=t'$ then we have the same
% equation as Eq.~(\ref{KE}) (`tilde' there should have been an
% overbar, according to the notation used in the notes from the very
% beginning to denote quantities in rotating frame, so I restore it to
% an overbar below):
%
\begin{equation}
\frac{\partial \hat{f}}{\partial T} +i[\hat{H}_0,\hat{f}] = 0.
\end{equation}

\subsection{Quasiparticle Lifetimes} \label{Sec:QPL}

We start from Eq.~(\ref{eq11}) in the main text
\begin{equation}
\hat{G}^R =
\frac{1}{\Lambda}\left[{\hat{G}_{0}^{R}-\det(\hat{G}_0^{R})\sigma_y
  (\hat{\Sigma}^R)^{\mathrm{T}}\sigma_y}\right].  \nonumber %\label{eq11}
\end{equation}
From Eqs.~(\ref{eq5})-(\ref{eq5_1}) it can be easily evaluated that
$\det(\hat{G}_0^{R}) = 1/(\varepsilon^2-\Omega^2+i\delta)$. Upon
substitution of Eq.~(\ref{eq5}), the first term of Eq.~(\ref{eq11}) can be written 
as follows 
\begin{widetext}
\begin{eqnarray}
\frac{\hat{G}_{0}^{R}}{\Lambda} 
  &=& \frac{\hat{A}}{\varepsilon-\Omega-\mathrm{Tr}(\hat{A}\hat{\Sigma}^R)-\mathrm{Tr}(\hat{B}\hat{\Sigma}^R)
  (\varepsilon-\Omega)/(\varepsilon+\Omega)+\det(\hat{\Sigma}^R)/(\varepsilon+\Omega)}
  \nonumber \\
&+&\frac{\hat{B}}{\varepsilon+\Omega-\mathrm{Tr}(\hat{A}\hat{\Sigma}^R)(\varepsilon+\Omega)/(\varepsilon-\Omega)-\mathrm{Tr}(\hat{B}\hat{\Sigma}^R)+\det(\hat{\Sigma}^R)/(\varepsilon-\Omega)}.  %\label{eq11}
\end{eqnarray}
\end{widetext}
In the vicinity of the poles we have
%\begin{widetext}
\begin{eqnarray}
\frac{\hat{G}_{0}^{R}}{\Lambda} 
  &\approx&
            \frac{\hat{A}}{\varepsilon-\Omega-\mathrm{Tr}(\hat{A}\hat{\Sigma}^R)\vert_{\varepsilon
            = \Omega} +\det(\hat{\Sigma}^R) \vert_{\varepsilon
            = \Omega}/(2\Omega)}
 \label{eqSB1} \\
&+&\frac{\hat{B}}{\varepsilon+\Omega-\mathrm{Tr}(\hat{B}\hat{\Sigma}^R)\vert_{\varepsilon
            = -\Omega}-\det(\hat{\Sigma}^R)\vert_{\varepsilon
            = -\Omega}/(2\Omega)}.  \nonumber 
\end{eqnarray}
%\end{widetext}
%
Since $\hat{\Sigma}^R \sim 1/\tau$ 
%Let $1/\tau$ denotes the order of magnitude of $\hat{\Sigma}^R$, 
where $\tau$ is the quasiparticle lifetime, $1/\tau \ll \Omega$ is
satisfied for our perturbative calculations. 
%For perturbative calculations, $1/\tau \ll \Omega$ is satisfied. 
We see that the last terms in the denominators above $\det(\hat{\Sigma}^R)\vert_{\varepsilon=\pm\Omega}/(2\Omega) \sim (1/\tau)\times 1/(\Omega\tau)$ 
are a factor of $1/(\Omega\tau)$ smaller than $1/\tau$ and thus can be neglected. Next we consider the second term of Eq.~(\ref{eq11}), whereupon substituting Eq.~(\ref{eq5})
        becomes 
\begin{widetext}
\begin{eqnarray}
\frac{\det(\hat{G}_0^{R})\sigma_y
  (\hat{\Sigma}^R)^{\mathrm{T}}\sigma_y}{\Lambda} &=& 
\frac{\sigma_y(\hat{\Sigma}^R)^{\mathrm{T}}\sigma_y}{\varepsilon^2-\Omega^2-(\varepsilon+\Omega)\mathrm{Tr}(\hat{A}\hat{\Sigma}^R)-(\varepsilon-\Omega)\mathrm{Tr}(\hat{B}\hat{\Sigma}^R)+\det(\hat{\Sigma}^R)}. \label{eqSB2} 
\end{eqnarray}
\end{widetext}
In the vicinity of the upper level where $\varepsilon \approx \Omega$, 
%
%\begin{widetext}
\begin{eqnarray}
&&\frac{\det(\hat{G}_0^{R})\sigma_y
  (\hat{\Sigma}^R)^{\mathrm{T}}\sigma_y}{\Lambda} \label{eqSB3}   \\
&\approx& \frac{1}{2\Omega}
\frac{\sigma_y(\hat{\Sigma}^R)^{\mathrm{T}}\sigma_y}{\varepsilon-\Omega-\mathrm{Tr}(\hat{A}\hat{\Sigma}^R)\vert_{\varepsilon
            = \Omega}+\det(\hat{\Sigma}^R)\vert_{\varepsilon
            = \Omega}/(2\Omega)}. \nonumber
%\nonumber  \\
%&\sim& \frac{1}{\Omega
%       \tau}\frac{1}{\varepsilon-\Omega-\mathrm{Tr}(\hat{A}\hat{\Sigma}^R)}. \label{eqSB3} 
\end{eqnarray}
%\end{widetext}
%
We see that the expression in the last line above is a factor of $1/(\Omega \tau)$
smaller than the corresponding $\varepsilon \approx \Omega$ 
contribution (\textit{i.e.}, the term $\propto \hat{A}$) in
Eq.~(\ref{eqSB1}), and hence can be neglected. A similar analysis
shows that the same is true for the $\varepsilon \approx -\Omega$ contribution in
Eq.~(\ref{eqSB2}). 
%In summary of all the analyses above, we have shown that effectively the terms proportional to $\det(\hat{G}_0^{R})$ in
%Eq.~(\ref{eq11}) are negligible when $1/(\Omega \tau) \ll 1$. 
Therefore we find
\begin{eqnarray}
\hat{G}^R(\epsilon) &\approx&  \frac{\hat{A}}{\varepsilon-\Omega-\mathrm{Tr}(\hat{A}\hat{\Sigma}^R)\vert_{\varepsilon
            = \Omega}} \nonumber \\
&+&\frac{\hat{B}}{\varepsilon+\Omega-\mathrm{Tr}(\hat{B}\hat{\Sigma}^R)\vert_{\varepsilon
            = -\Omega}}, 
\end{eqnarray}
which gives Eq.~(\ref{eq12}) with Eqs.~(\ref{eq13_1})-(\ref{eq13_2}) in the main text.

\subsection{TLS Self-Energy} \label{Sec:TLS_SE}

Here we derive the expression for the TLS self-energy. Analytic expression of the diagram depicted in Fig.~\ref{Figure1} reads
\begin{gather}\label{A1}\nonumber
\hat{\Sigma}(t-t')=i\int \frac{d\textbf{k}}{(2\pi)^2}\hat{M}_{\textbf{k}}\hat{{G}}(t-t')\hat{M}_{-\textbf{k}}\Pi(\textbf{k},t-t'),\\
\Pi(\textbf{k},t-t')=i\sum_{\textbf{p}}\mathcal{G}(\textbf{p},t-t')\mathcal{G}(\textbf{p}+\textbf{k},t'-t),
\end{gather}
where times $t,t'$ are located on the Keldysh contour, $\Pi$ and
$\mathcal{G}$ are the polarization operator and Green's function of
the bath's particles.

To proceed further, let us first perform an analytic continuation to the real time domain. Using the Langreth's rules~\cite{14_Jauhobook} we find
\begin{eqnarray}
\hat{\Sigma}^<(\omega) &=&
   i\sum_{\textbf{k},\varepsilon}\hat{M}_{\textbf{k}}\hat{{G}}^<(\omega-\varepsilon)\hat{M}_{-\textbf{k}}\Pi^<(\textbf{k},\varepsilon),
\label{App2_1} \\
\hat{\Sigma}^R(\omega) &=&
                           i\sum_{\textbf{k},\varepsilon}\hat{M}_{\textbf{k}}\left[\hat{{G}}^<(\omega-\varepsilon)\Pi^R(\textbf{k},\varepsilon) \right.\nonumber \\
&&+\hat{{G}}^R(\omega-\varepsilon)\Pi^<(\textbf{k},\varepsilon) \nonumber\\
&&\left.+\hat{{G}}^R(\omega-\varepsilon)\Pi^R(\textbf{k},\varepsilon)\right]\hat{M}_{-\textbf{k}},
\label{App2_2}
\end{eqnarray}
\begin{eqnarray}
\Pi^<(\textbf{k},\omega) &=& i\sum_{\textbf{p},\varepsilon}\mathcal{G}^<(\textbf{p}+\textbf{k},\varepsilon+\omega)\mathcal{G}^>(\textbf{p},\varepsilon),\label{App2_3} \\
\Pi^R(\textbf{k},\omega) &=&
   i\sum_{\textbf{p},\varepsilon}[\mathcal{G}^<(\textbf{p}+\textbf{k},\varepsilon+\omega)\mathcal{G}^A(\textbf{p},\varepsilon)
  \nonumber \\
&&+\mathcal{G}^R(\textbf{p}+\textbf{k},\varepsilon+\omega)\mathcal{G}^<(\textbf{p},\varepsilon)].\label{App2_4}
\end{eqnarray}
%
% \begin{gather}\label{App2_1}
% \hat{\Sigma}^<(\omega)=i\sum_{\textbf{k},\epsilon}\hat{M}_{\textbf{k}}\hat{{G}}^<(\omega-\epsilon)\hat{M}_{-\textbf{k}}\Pi^<(\textbf{k},\epsilon);\\\nonumber
% \hat{\Sigma}^R(\omega)=i\sum_{\textbf{k},\epsilon}\hat{M}_{\textbf{k}}[\hat{{G}}^<(\omega-\epsilon)\Pi^R(\textbf{k},\epsilon)+\\\nonumber
% +\hat{{G}}^R(\omega-\epsilon)\Pi^<(\textbf{k},\epsilon)+
% \hat{{G}}^R(\omega-\epsilon)\Pi^R(\textbf{k},\epsilon)]\hat{M}_{-\textbf{k}};\\\nonumber
% \Pi^<(\textbf{k},\omega)=i\sum_{\textbf{p},\epsilon}\mathcal{G}^<(\textbf{p}+\textbf{k},\epsilon+\omega)\mathcal{G}^>(\textbf{p},\epsilon);\\\nonumber
% \Pi^R(\textbf{k},\omega)=i\sum_{\textbf{p},\epsilon}[\mathcal{G}^<(\textbf{p}+\textbf{k},\epsilon+\omega)\mathcal{G}^A(\textbf{p},\epsilon)+\\
% +\mathcal{G}^R(\textbf{p}+\textbf{k},\epsilon+\omega)\mathcal{G}^<(\textbf{p},\epsilon)]\nonumber
% \end{gather}
%
\subsubsection{TLS self-energy in normal state bath}
Lets consider the bath in normal phase state, $T>T_c$. In this case
the bare Green's functions of the bath's particles are
%n_B(\xi_{\textbf{p}})
\begin{eqnarray}%\label{A3}
\mathcal{G}^<(\textbf{k},\varepsilon) &=&-2\pi i
                                       n_B(\xi_{\textbf{k}})\delta(\varepsilon-E_{\textbf{k}}),
  \label{A3_1}\\
\mathcal{G}^>(\textbf{k},\varepsilon)&=&-2\pi i[1+n_B(\xi_{\textbf{k}})]\delta(\varepsilon-E_{\textbf{k}}),\label{A3_2}\\
\mathcal{G}^R(\textbf{k},\varepsilon)&=&\frac{1}{\varepsilon-E_{\textbf{k}}+i\delta},\label{A3_3}
%\mathcal{G}^<(\textbf{k},\epsilon)=-2\pi in_{\textbf{k}}\delta(\epsilon-E_{\textbf{k}})\\\nonumber
%\mathcal{G}^>(\textbf{k},\epsilon)=-2\pi i[n_{\textbf{k}}+1]\delta(\epsilon-E_{\textbf{k}})\\\nonumber
%\mathcal{G}^R(\textbf{k},\epsilon)=\frac{1}{\epsilon-E_{\textbf{k}}+i\delta}
\end{eqnarray}
where $n_B(\xi_{\textbf{k}})$ is equilibrium Bose distribution and $E_{\textbf{k}}=k^2/2m$ is the energy of the bath's particles (which
corresponds to the kinetic energy of the exciton's center-of-mass
motion for the excitonic bath we consider in Section \ref{Sec:QD}.
Using these functions we find the following polarization operators
\begin{eqnarray}%\label{A4}\nonumber
\Pi^<(\textbf{k},\omega)&=&-2\pi
                            i\sum_{\textbf{p}}n_B(\xi_{\textbf{k}+\textbf{p}})[1+n_B(\xi_{\textbf{p}})]
                            \nonumber \\
&&\times\delta(\omega+E_{\textbf{p}}-E_{\textbf{p}+\textbf{k}}), \label{A4_1}\\
\Pi^R(\textbf{k},\omega)&=&\sum_{\textbf{p}}\frac{n_B(\xi_{\textbf{p}})-n_B(\xi_{\textbf{k}+\textbf{p}})}{\omega+E_{\textbf{p}}-E_{\textbf{p}+\textbf{k}}+i\delta}. \label{A4_2}
%\Pi^<(\textbf{k},\omega)=-2\pi i\sum_{\textbf{p}}n_{\textbf{p}+\textbf{k}}(n_{\textbf{p}}+1)\delta(\omega+E_{\textbf{p}}-E_{\textbf{p}+\textbf{k}})\\
%\Pi^R(\textbf{k},\omega)=\sum_{\textbf{p}}\frac{n_{\textbf{p}}-n_{\textbf{p}+\textbf{k}}}{\omega+E_{\textbf{p}}-E_{\textbf{p}+\textbf{k}}+i\delta}
\end{eqnarray}
Substituting these expressions into Eq.~(\ref{App2_1}), we obtain the self-energy expression Eq.~(\ref{eq14}).
\subsubsection{TLS self-energy in BEC bath}
If the bath is in the Bose-condensed state, the elementary excitations
in the Bogoliubov's theory of weakly-interacting Bose gas have
the energy dispersion
\begin{equation}
\epsilon_{\textbf{k}}=\sqrt{\frac{k^2}{2m}\left(\frac{k^2}{2m}+2ms^2\right)}, \label{A7_0}
\end{equation}
where $s^2=g_0n_c/m$ is the Bogolubov quasiparticle's speed of sound,
$g_0$ is the inter-particle interaction strength, and $n_c$ particles
density in the condensate. The Green functions are given by
\begin{eqnarray}\label{A7}
\hat{\mathfrak{G}}^R(\textbf{k},\varepsilon) &=& \left(
                                          \begin{array}{cc}
                                            \mathfrak{G} & \mathfrak{F}^+ \\
                                            \mathfrak{F} & \tilde{\mathfrak{G}} \\
                                          \end{array}
                                        \right)^R=\frac{1}{(\varepsilon+i\delta)^2-\epsilon_{\textbf{k}}^2}\label{A7_1}\\
&&\times\left(
\begin{array}{cc}
  \varepsilon+k^2/2m+ms^2 & -ms^2 \\
  -ms^2 & -\varepsilon+k^2/2m+ms^2 \\
\end{array}\right),
\nonumber
\end{eqnarray}
and
%n_B(\xi_{\textbf{p}})
\begin{eqnarray}%\label{A8}
&&\hat{\mathfrak{G}}^<(\textbf{k},\varepsilon) =
                                              n_B(\varepsilon)[\hat{\mathfrak{G}}^R(\textbf{k},\varepsilon)-\hat{\mathfrak{G}}^A(\textbf{k},\varepsilon)]
                                              \nonumber \\
&=& -\frac{2\pi i}{2\epsilon_{\textbf{k}}}\left(
\begin{array}{cc}
  \varepsilon+k^2/2m+ms^2 & -ms^2 \\
  -ms^2 & -\varepsilon+k^2/2m+ms^2 \\
\end{array}\right)\nonumber \\
&&\times\left\{n_B(\epsilon_{\textbf{k}})\delta(\varepsilon-\epsilon_{\textbf{k}})+[1+n_B(\epsilon_{\textbf{k}})]\delta(\varepsilon+\epsilon_{\textbf{k}})\right\},\label{A8_1}
\end{eqnarray}
\begin{eqnarray}
&&\hat{\mathfrak{G}}^>(\textbf{k},\varepsilon) = [1+n_B(\varepsilon)][\hat{\mathfrak{G}}^R(\textbf{k},\varepsilon)-\hat{\mathfrak{G}}^A(\textbf{k},\varepsilon)]\nonumber
  \\
&=&-\frac{2\pi i}{2\epsilon_{\textbf{k}}}\left(
\begin{array}{cc}
  \varepsilon+k^2/2m+ms^2 & -ms^2 \\
  -ms^2 & -\varepsilon+k^2/2m+ms^2 \\
\end{array}\right)\nonumber \\
&&\times\left\{[1+n_B(\epsilon_{\textbf{k}})]\delta(\varepsilon-\epsilon_{\textbf{k}})+n_B(\epsilon_{\textbf{k}})\delta(\varepsilon+\epsilon_{\textbf{k}})\right\}.\label{A8_2}
% \hat{\mathfrak{G}}^<(\textbf{k},\epsilon)=N_\epsilon[\hat{\mathfrak{G}}^R(\textbf{k},\epsilon)-\hat{\mathfrak{G}}^A(\textbf{k},\epsilon)]\\\nonumber=-\frac{2\pi i}{2\epsilon_k}\left(
% \begin{array}{cc}
%   \epsilon+k^2/2m+ms^2 & -ms^2 \\
%   -ms^2 & -\epsilon+k^2/2m+ms^2 \\
% \end{array}\right)\times\\\nonumber
% \times[N_k\delta(\epsilon-\epsilon_k)+(1+N_k)\delta(\epsilon+\epsilon_k)],\\\nonumber
% \hat{\mathfrak{G}}^>(\textbf{k},\epsilon)=(1+N_\epsilon)[\hat{\mathfrak{G}}^R(\textbf{k},\epsilon)-\hat{\mathfrak{G}}^A(\textbf{k},\epsilon)]\\\nonumber=-\frac{2\pi i}{2\epsilon_k}\left(
% \begin{array}{cc}
%   \epsilon+k^2/2m+ms^2 & -ms^2 \\
%   -ms^2 & -\epsilon+k^2/2m+ms^2 \\
% \end{array}\right)\times\\\nonumber
% \times[(1+N_k)\delta(\epsilon-\epsilon_k)+N_k\delta(\epsilon+\epsilon_k)],\\\nonumber
% N_k=\frac{1}{e^{\epsilon_k/T}-1},
\end{eqnarray}
The polarization operator has two contributions. The first one comes
from the condensate particles and the other one from non-condensate
particles, Fig.~\ref{Figure1}. If the Bose bath is a two-dimensional
system, the condensate occurs at zero temperature only. In this case
one assumes that $n_B(\epsilon_{\textbf{k}})=0$. The contribution
$P_c(\textbf{k},\omega)$ to the polarization operator from the condensate particles is
\begin{eqnarray} %\label{A9}\nonumber
P^R_c(\textbf{k},\omega) &=&
                             n_c[\mathfrak{G}^R+\mathfrak{F}^{+R}+\mathfrak{F}^R+\tilde{\mathfrak{G}}^R] \nonumber \\
&=& n_c\frac{k^2/m}{(\omega+i\delta)^2-\epsilon_{\textbf{k}}^2}, \label{A9_1}\\
P^<_c(\textbf{k},\omega) &=&
                             n_c[\mathfrak{G}^<+\mathfrak{F}^{+<}+\mathfrak{F}^<+\tilde{\mathfrak{G}}^<]
                             \nonumber \\
&=&-\pi in_c\frac{k^2/m}{\epsilon_{\textbf{k}}}\delta(\omega+\epsilon_{\textbf{k}}).\label{A9_2}
% P^R_c(\textbf{k},\omega) &=&
%                              n_c[\mathfrak{G}^R+\mathfrak{F}^{+R}+\mathfrak{F}^R+\tilde{\mathfrak{G}}^R] \nonumber \\
% &=& n_c\frac{k^2/m}{(\varepsilon+i\delta)^2-\epsilon_{\textbf{k}}^2}, \label{A9_1}\\
% P^<_c(\textbf{k},\omega) &=&
%                              n_c[\mathfrak{G}^<+\mathfrak{F}^{+<}+\mathfrak{F}^<+\tilde{\mathfrak{G}}^<]
%                              \nonumber \\
% &=&-\pi in_c\frac{k^2/m}{\epsilon_k}\delta(\varepsilon+\epsilon_{\textbf{k}}).\label{A9_2}
% %P^R_c(\textbf{k},\omega)=n_c[\mathfrak{G}^R+\mathfrak{F}^{+R}+\mathfrak{F}^R+\tilde{\mathfrak{G}}^R]=n_c\frac{k^2/m}{(\epsilon+i\delta)^2-\epsilon_k^2},\\\nonumber
% %P^<_c(\textbf{k},\omega)=n_c[\mathfrak{G}^<+\mathfrak{F}^{+<}+\mathfrak{F}^<+\tilde{\mathfrak{G}}^<]\\=-2\pi in_c\frac{k^2/m}{2\epsilon_k}\delta(\epsilon+\epsilon_k).
\end{eqnarray}
Using this functions one finds the condensate contribution to the retarded self-energy, Eq.~(\ref{eq16_1}).

Now let us find the contribution to the self-energy from the non-condensate particles.
First, we need the retarded and lesser polarization operators. In the
regime $ms^2 \gg k^2/2m$ where linear dispersion
$\epsilon_{\textbf{k}}=sk$ of the Bogoliubov quasiparticles holds,
using Eqs.~(\ref{A7})-(\ref{A8_2}) we find for the retarded and lesser polarization operators of non-condensate particles
\begin{eqnarray}%\label{A11}
P_n^R(\textbf{k},\omega) &=&
2i\sum_{\textbf{p},\varepsilon}\left[\mathfrak{F}^<(\textbf{p}+\textbf{k},\varepsilon+\omega)\mathfrak{F}^{+A}(\textbf{p},\varepsilon)\right.\label{A11_1}
\\
&&\left.+\mathfrak{F}^R(\textbf{p}+\textbf{k},\varepsilon+\omega)\mathfrak{F}^{+<}(\textbf{p},\varepsilon)\right]\nonumber \\
&=&\frac{(ms^2)^2}{2}\sum_{\textbf{p}}\frac{1}{\epsilon_{\textbf{p}}\epsilon_{\textbf{p}+\textbf{k}}}\left(\frac{1}{\omega-\epsilon_{\textbf{p}}-\epsilon_{\textbf{p}+\textbf{k}}+i\delta}\right. \nonumber\\
&&\left.-\frac{1}{\omega+\epsilon_{\textbf{p}}+\epsilon_{\textbf{p}+\textbf{k}}+i\delta}\right),\nonumber
\end{eqnarray}
\begin{eqnarray}
P_n^<(\textbf{k},\omega) &=&
                             2i\sum_{\textbf{p},\varepsilon}\mathfrak{F}^<(\textbf{p}+\textbf{k},\varepsilon+\omega)\mathfrak{F}^{+>}(\textbf{p},\varepsilon)\label{A11_2} \\
&=&-\pi
    i(ms^2)^2\sum_{\textbf{p}}\frac{1}{\epsilon_{\textbf{p}}\epsilon_{\textbf{p}+\textbf{k}}}\delta(\omega+\epsilon_{\textbf{p}}+\epsilon_{\textbf{p}+\textbf{k}}). \nonumber
% P_n^R(\textbf{k},\omega)=2i\sum_{\textbf{p},\epsilon}[\mathfrak{F}^<(\textbf{p}+\textbf{k},\epsilon+\omega)\mathfrak{F}^{+A}(\textbf{p},\epsilon)+\\\nonumber
% +\mathfrak{F}^R(\textbf{p}+\textbf{k},\epsilon+\omega)\mathfrak{F}^{+<}(\textbf{p},\epsilon)]\\\nonumber
% =\frac{(Ms^2)^2}{2}\sum_{\textbf{p}}\frac{1}{\epsilon_{\textbf{p}}\epsilon_{\textbf{p}+\textbf{k}}}\left[\frac{1}{\omega-\epsilon_{\textbf{p}}-\epsilon_{\textbf{p}+\textbf{k}}+i\delta}-\frac{1}{\omega+\epsilon_{\textbf{p}}+\epsilon_{\textbf{p}+\textbf{k}}+i\delta}\right],\\\nonumber
% P_n^<(\textbf{k},\omega)=2i\sum_{\textbf{p},\epsilon}\mathfrak{F}^<(\textbf{p}+\textbf{k},\epsilon+\omega)\mathfrak{F}^{+>}(\textbf{p},\epsilon)\\\nonumber
% =-2\pi i\frac{(Ms^2)^2}{2}\sum_{\textbf{p}}\frac{1}{\epsilon_{\textbf{p}}\epsilon_{\textbf{p}+\textbf{k}}}\delta(\omega+\epsilon_{\textbf{p}}+\epsilon_{\textbf{p}+\textbf{k}}).
\end{eqnarray}
Now the calculation of the non-condensate particles' contribution to
the self-energy of TLS is simple, and we arrive at the expression Eq.~(\ref{eq16_2}) of the main text.


\begin{thebibliography}{50}
\bibitem{Remark1} This follows from our assumption of
  low-temperatures: the relaxation rate in the normal state obtained
  in the limit of low temperatures $T \ll \Omega$ is a
  zero-temperature result, while the relaxation rate in the
  Bose-condensed phase is obtained for temperatures below the BEC critical
  temperature (typically a few kelvins) where the Bogoliubov theory at
  $T = 0$ applies. At higher temperatures, the relaxation rates become
  non-zero when $\omega < 2\Delta$, as indicated by our result
  Eq.~(\ref{eq22}).
\end{thebibliography}
\end{document}